\documentclass[sigconf, s]{acmart}
\usepackage{algpseudocode}
\usepackage{algorithm}
\usepackage{cleveref}
\usepackage{caption}
\usepackage{pifont}
\usepackage[colorinlistoftodos,prependcaption,textsize=tiny]{todonotes}

\newcommand{\cmark}{\ding{51}}%
\newcommand{\xmark}{\ding{55}}%
\usepackage{subcaption}
\algnewcommand\algorithmicinput{\textbf{Input:}}
\algnewcommand\algorithmicoutput{\textbf{Output:}}
\algnewcommand\Input{\item[\algorithmicinput]}%
\algnewcommand\Output{\item[\algorithmicoutput]}%

\AtBeginDocument{%
  \providecommand\BibTeX{{%
    \normalfont B\kern-0.5em{\scshape i\kern-0.25em b}\kern-0.8em\TeX}}}

\acmConference[GTX Technical Report]{Online}{2025}{arxiv}
\acmPrice{15.00}
\newcommand{\sys}{GTX}
\begin{document}

%\section*{Miscellaneous Modifications}
%We have added a few minor modifications to fix typos and to explain things better. We fixed the typo in Figure~\ref{fig:edge-deltas-block-example}. We added a 
%line 
%sentence
%in Section~\ref{sec-mixed-workload} to 
%that
%indicates that all LiveGraph experiments are conducted using only 20\% of the logs. We also fixed a typo in Section~\ref{sec-mixed-workload}; we used the word "check" but it should be "checked".
\title{
\sys{}: A  Write-Optimized Latch-free Graph Data System with Transactional Support - Extended Version\\
%can also be GTL or LF-GTX (lock-free graph system with transactional support) 
%\walid{Let me know what you think of the name.}
%supports efficient transactions and purely sequential adjacency list scan
%\Libin{I like the name, I think LFGTX may be better, we can rebrand as latch or lock free here. But if GTX is more concise I feel. And looks cooler and simpler.}
%\walid{Do we change the title to be latch free or leave it as lock free?}
}
%\subtitle{(Extended Version)}

%\author{Libin Zhou,~~Lu Xing, ~~Yeasir Rayhan,  and~~Walid G. Aref}
%\email{{zhou822, aref, xingl, yrayhan}@purdue.edu}
%\affiliation{%
  %\institution{Purdue University, West Lafayette, Indiana, USA}
%}

\author{Libin Zhou, Lu Xing, Yeasir Rayhan, Walid. G. Aref}
\affiliation{%
  \institution{Purdue University}
  \streetaddress{610 Purdue Mall}
  \city{West Lafayette}
  \state{Indiana}
  \country{USA}
  \postcode{47907}
  }
\email{{zhou822,xingl,yrayhan,aref}@purdue.edu}


\begin{abstract}
%This paper introduces \sys, a standalone main-memory write-optimized graph data system that specializes in structural and graph property updates while enabling concurrent reads and graph analytics through ACID transactions. Recent graph systems target concurrent read and write support while guaranteeing transaction semantics. However, their performance suffers from updates with real-world temporal locality over the same vertices and edges due to vertex-centric lock contentions. \sys~ introduces an adaptive delta chain locking protocol on top of a carefully designed latch-free graph storage. It eliminates vertex-level locking contention, and adapts to real-life workloads while maintaining sequential access to the graph's adjacency lists storage. \sys's transactions further support fast group commit and cooperative garbage collection. This combination of features ensures high update throughput and provides low latency graph analytics.Based on experimental evaluation, in addition to maintaining competitive concurrent read and analytical performance, \sys~ has high read-write transaction throughput. For write-heavy transactional workloads, \sys{} achieves up to $11\times$ better transaction throughput than the best-performing state-of-the-art system. At the same time, \sys{} does not sacrifice the performance of read-heavy analytical workloads, and has competitive performance similar to state-of-the-art systems. 
This paper introduces \sys, a standalone main-memory write-optimized graph data system that specializes in structural and graph property updates while enabling concurrent reads and graph analytics through ACID transactions. Recent graph systems target concurrent read and write support while guaranteeing transaction semantics. However, their performance suffers from updates with real-world temporal locality over the same vertices and edges due to vertex-centric lock contentions. \sys~ has an adaptive delta-chain locking protocol on top of a carefully designed latch-free graph storage.
It eliminates vertex-level locking contention, and adapts to real-life workloads while maintaining sequential access to the graph's adjacency lists storage. \sys's transactions further support cache-friendly block-level concurrency control, and cooperative group commit and garbage collection. This combination of features ensures high update throughput and provides low latency graph analytics. 
Based on experimental evaluation, in addition to not sacrificing the performance of read-heavy analytical workloads, and having competitive performance similar to state-of-the-art systems, \sys~ has high read-write transaction throughput. For write-heavy transactional workloads, \sys{} achieves 
up to $11\times$ better transaction throughput than the best-performing state-of-the-art system.

%\wga{I wrote a sample above. Try to edit it according to the actual performance you got.}
%\Libin{March 12th: add some real number}
%\walid{may want to add one performance number here, eg outperforms state of the art system by up to xxx times}
%\Libin{March 5th: I have multiple numbers, which one should I put?}
%\Libin{March 5th, changed some wording regarding the locks and latches, be more specific, change "locks" to "latches". With better experiment results we can claim we achieve good mixed workload result}
%Moreover, \sys~ maintains  
%while maintaining 
%competitive concurrent read and analytical performance.
\end{abstract}


%Libin: I could not find much better CCS to be there
%\ccsdesc[100]{Information systems~Data structures}

\settopmatter{printacmref=false} % Removes citation information below abstract
\maketitle

\section{ Introduction}
\label{sec-intro}
%\walid{change the format to sigmod format and not vldb.}
%Libin: should be done now
%Libin: Dec. 29th: Highly Conuccrent Update Efficient Graph: HCUE(G) SPIDER
%Libin: Jan 10th: add our contribution
%\Libin{April 13th: add vspace to reduce space between sections and between figures}
%\wga{Please do not use vspace anywhere. I have a better solution.}
Managing and querying dynamic graphs have become increasingly important in many real-world applications, 
%like 
e.g.,
%WGA: no social network?
recommendation systems, fraud detection, threat detection, geo-spatial navigation, e-commerce, knowledge graph applications, and risk management~\cite{twitter-recommendation-ref2, twitter-recoomendation-ref, dynamic-spatial-ref, graph-challenge-ref, byte-graph-ref,authentication-graph-ref,sap-hana-graph,maria-graph,oracle-graph}. Graphs continuously change~\cite{social-network-ref,linkbench-ref,alibab-talk-ref,content-rec-ref,byte-graph-ref,real-time-cycle-ref,china-shopping-ref,tweets-ref,dynamic-graph-streaming-ref,large-dynamic-graph-ref} with millions of edge updates per second~\cite{dynamic-graph-streaming-ref}, e.g., 500 million tweets generated in Twitter per day~\cite{large-dynamic-graph-ref}.
%reaching millions of updates per second, e.g.,  in social networks~\cite{alibab-talk-ref,content-rec-ref,byte-graph-ref,real-time-cycle-ref,china-shopping-ref,tweets-ref}. 
Applications 
%also 
require transactions
%al atomicity 
to update multiple vertices and edges~\cite{byte-graph-ref}, e.g., in ByteDance services~\cite{bytedance-ref}, 
%\walid{Add reference for bytedance here.}
when a user creates an article, a transaction atomically inserts  3 edges: (user, article), (user, tag), (article, tag). 
%\Libin{the following are added on Feb. 19th to further address the need of transactions} 
Without transactions, updates can corrupt the data~\cite{chengmammoths}, 
%e.g. 
e.g.,
violating the reciprocal consistency that requires atomically updating two edges between a pair of vertices~\cite{chengmammoths,rec-consis-ref}.
At the same time, dynamic graphs need to support graph analytics~\cite{Sortledton, Teseo, LiveGraph, byte-graph-ref,bytegap-ref}. ByteDance~\cite{bytedance-ref} detects fraud, and manages risk via subgraph pattern-matching that may traverse multiple hops while the graph is being updated concurrently~\cite{byte-graph-ref}. Without transactional atomicity, consistency, and isolation,  graph analytics algorithms, originally designed for static graphs, cannot be run~\cite{Sortledton}, and may yield incorrect results~\cite{Teseo,chengmammoths} and 
security vulnerabilities~\cite{chengmammoths,concurrency-attacks-web-ref},
%In fraud detection, a customer may be  flagged wrongly. In computer networks, a suspicious authentication may pass  security checks without being noticed~\cite{Teseo}. Also, 
e.g., an access control system needs to update access permissions of roles and entities atomically~\cite{chengmammoths,concurrency-attacks-web-ref}. Without transactional guarantees, users may obtain incorrect permissions, and malicious users can exploit race conditions to trigger vulnerabilities. %Although applications can build concurrency control protocols to achieve transactional isolation, these solutions are inefficient and complicated~\cite{chengmammoths}.
%\Libin{end of added ones on Feb. 19th}
% yeasir: May be name those systems explicitly, here and in the next line as well. Libin: if we have space I will do that!
Graph systems, e.g.,~\cite{llama-ref,graph1-ref,stinger-ref}, 
%don't 
that do not
support transactions cannot run graph analytics concurrently with updates~\cite{Sortledton}. Recent graph data systems, e.g.,~\cite{LiveGraph,Sortledton,Teseo}, support mixed transactions and analytics workloads.

% yeasir: Can we shift the previous line here and then continue the paragraph saying what is the problem with the SOTA systems like you have here.
Experiments~\cite{Sortledton} show that 
% yeasir: include 'the' before performance?
the performance of state-of-the-art transactional graph systems~\cite{LiveGraph,Teseo,Sortledton} 
% yeasir: a typo? = 'suffer' Libin: fixed
suffer significantly %given write-heavy workloads 
when workloads follow real-world update patterns. Many real-world scenarios exhibit power-law graphs with vertex degrees following power-law distribution~\cite{bank-power-law,powerlyra-ref,power-law-ref,LiveGraph,taobench-ref}, and super vertices (hub vertices)~\cite{byte-graph-ref, Sortledton}) having 
% include a = a large fanout
large degree. Also, real-world graph workloads have hotspots, and their edge updates have temporal locality~\cite{konect-ref,Sortledton}. These update patterns cause not only updates to the same vertex's adjacency list to be temporally close to each other, 
but also create congestion of large amounts of concurrent updates at the same vertices (i.e., hub vertices becoming hotspots~\cite{byte-graph-ref}). Vertex-centric locking and the lack of lock-free synchronization 
cause significant degradation in %performance, especially in terms of 
transaction throughput~\cite{Sortledton}. 
%\walid{Is the above addition correct and relevant?}
%\Libin{I think it is relevant because it addresses the current challenges faced by SOTR systems and what are believed to be the bottlenecks/causes. But what do you mean by "correct here"?}

%\Libin{April 11th: how about we also add a phrase saying our storage is latch free?} 
%\wga{Add it in the contributions bullet list.}
% yeasir: while reading, latch-free indexes pop up suddenly. Maybe say something like ; Even though there exists latch-free indexes to get around the problem of locking -- just add a connector so that it connects with the previous paragraph
%Libin: I agree, I have added "To eliminate locks' blocking," as a connector.
%To eliminate locks' blocking, 
Latch-free indexes, e.g.,~\cite{BwTree, openBwTree, bztree-ref, quadboost-ref, lock-free-data-structure-ref, lock-free-slides-ref, EPVS, Deuteronomy-ref}, eliminate locks and blocking, and hence offer higher concurrency.
%\lu{I think latch need to be capitalized.}
They rely on atomic hardware primitives, e.g.,
%like 
{\em compare\_and\_swap (CAS)} and 
{\em fetch\_add}   to serialize updates, but by themselves, do not support transactions~\cite{BwTree}. %the notion of transactions, but serve as building blocks to achieve that~\cite{BwTree}.
%For example, 
%The Bw-Tree~\cite{BwTree, openBwTree}, has a separate transaction protocol~\cite{hekaton-ref} to handle transactions. Thus, it does not prevent write-write conflicts or provide transaction atomicity and consistency. 
The Bw-Tree~\cite{BwTree, openBwTree} does not prevent write-write conflicts or provide transaction atomicity and consistency. It requires a separate transaction protocol~\cite{hekaton-ref} to handle transactions.
%Extending a latch-free index to be a standalone system requires designing an efficient transaction protocol.

%Inspired by the latch-free indexes concurrent read-write thread execution, t
This paper studies realizing a high-throughput transactional graph data system, \sys{\footnote{The source code can be found at: https://github.com/Jiboxiake/GTX-SIGMOD2025}}, using a 
%atomic primitives-based 
latch-free graph store and multiversion concurrency control $MVCC$~\cite{mvcc-ref}. 
\sys{} is a latch-free write-optimized main-memory transactional graph system  supporting highly concurrent read-write transactions and graph analytics.

\sys{} has a cache-friendly latch-free $MVCC$ dynamic graph storage managed and accessed completely by non-blocking atomic primitives.
It combines pointer-based and sequential delta-chain storage for efficient edge lookup and adjacency list scans. It uses vector-based delta-chains indexes to locate target edges efficiently while preserving cache-effective sequential adjacency lists. \sys{}'s storage is optimized for power-law edge distributions and hub vertices that are prominent in real-world graphs~\cite{LiveGraph,Sortledton}. 
\sys{} has an efficient transaction protocol for the underlying latch-free data structure and dynamic graph workload. 
\sys{} eliminates vertex- and edge-centric locks, and has an adaptive delta-chain locking protocol. %write-write conflict prevention protocol. 
\sys{}
adapts to the temporal locality and hotspots of edge updates by increasing concurrency for ``hot" %adjacency lists 
vertices
with higher memory cost, and reducing concurrency for the ``cold" %blocks. 
vertices.
To exploit the high concurrency provided by latch-free $MVCC$ and adaptive concurrency control, \sys~ has a low-latency hybrid group commit protocol and %cache-friendly lock-free 
cooperative $MVCC$ garbage collection. \sys~ 
%is  efficient in computing resources with 
exhibits a cooperative worker thread design. Except for a single commit manager thread, \sys~ has no  service threads. \sys~  worker threads install committed transactions' updates, and collect
garbage 
%failed transaction aborts, and commit other transactions' updates 
while executing each thread's own transactions.

%Experiments demonstrate that 
%\lu{All the multiplication sign should be $\times$}
%\lu{Multiplication is like 2$\times$, not $2x$. Same goes for in the abstract.}
%Experiments demonstrate that \sys~ achieves up to $2\times$ higher transaction throughput in random order power-law graph edge insertions, and up to $11\times$ higher transaction throughput in real-world timestamp-ordered power-law graph edge insertions than the best competitor. 
Experiments demonstrate that \sys~ achieves up to $2\times$ and $11\times$ higher transaction throughput than the best competitor in random order power-law graph edge insertions and in real-world timestamp-ordered power-law graph edge insertions, respectively. 
%, and 1.2x higher transaction throughput in power-law graph edge updates than the best competitor. 
%\Libin{maybe we can remove the last "1.2x" part because we did not present this experiment in the evaluation section? Also changed the below sentence}
%For update transactions with temporal localities and graph analytics workloads, 
For concurrent transaction and graph analytics workloads, 
\sys~ has up to {\color{black}$5.3\times$ }%$4.3\times$ 
higher throughput in edge update transactions for write-heavy workloads, and {\color{black}$3.7\times$ }%$2.8\times$ 
higher throughput for read-write balanced workloads over {\color{black}the best} competitor systems.
The trade-off is that \sys~ takes between {\color{black}$0.95\times$ to $2\times$ }%$0.8\times$ to $1.8\times$ 
longer time than the best performing system to execute most of the concurrent graph analytics, but still performs reasonably well for the majority of the workloads. 
%are only 57\% and 88\% slower respectively. 
%\walid{Why did you say: "are only 57\% and 88\% slower respectively."? Are you comparing against multiple systems? You have not stated that here.}
%\Libin{I originally meant to save our write throughput is 150 to 300 percent better while read is only 60 percent slower, so our strength outperforms weakness. I want to  emphasize that we did not sacrifice too much of read.}
%In fact, \sys~ is one of the only two transactional graph systems that can handle mixed-workload without any errors. 

%\walid{If these bullets will not be strong, then, we can omit the whole paragraph. Try to copy from the bullets in our review response.}
%\Libin{I will try to make them stronger, I think I can. I just wrote some ideas down first}
%\Libin{I delete some detailed experiment results because one reviewer thinks we are too redundant}
The contributions of this paper are as follows.
%{\bf 1.}
We introduce \sys{}, a write-optimized transactional graph system with high transaction throughput.
%{\bf 2.}
\sys{} has a latch-free graph storage 
that leverages existing techniques of latch-free indexes, and is optimized for adjacency list storage of dynamic graphs. 
%{\bf 3.}
\sys{} has a delta chains-based adaptive $MVCC$, combining pointer-based delta-chains storage and sequential storage to provide both efficient single edge lookup and adjacency list scans.
%{\bf 4.}
\sys{} has a high-throughput transaction protocol equipped with a hybrid commit protocol, and %decentralized epoch-based 
cooperative garbage collection suited for latch-free multi-versioning storage. 
%{\bf 5.}
\sys{} concurrency control dynamically adapts to real-world update patterns, temporal localities, and hotspots, %For power-law graphs without temporal localities and hotspot, \sys{} transactions achieve up to 2.27 higher read-write transaction throughput than its competitors.
%\sys{} has a delta chains-based adaptive $MVCC$, 
and eliminates the need for vertex- or edge-centric locking commonly used in graph systems. 
%, where \sys~ achieves up to 11x better read-write transaction throughput for these workloads.
%\noindent
%{\bf 6.} 
%For mixed-workloads of read-write transactions and graph analytics, 
%\sys~ achieves high read-write transactional throughput while maintaining competitive graph analytics latency concurrently. 
%{\bf 6.}
Our 
%comprehensive 
experiments demonstrate that \sys~ achieves up to 11x higher read-write transaction throughput while maintaining competitive graph analytics latency concurrently.

The rest of this paper proceeds as follows. Section~\ref{sec-related-work} discusses related work on dynamic graph systems. Section~\ref{gtx-overview} overviews \sys{}. Sections~\ref{sec-graph-storage}
%, \ref{sec:txn-operations}, 
- \ref{sec-resource-management} present \sys{'s} graph store, transaction operations and transaction management. Section~\ref{sec-experiment} presents an extensive experimental study. Finally, Section~\ref{bwgraph-conclusion} concludes the paper.
\section{Related Work}%\Libin{we should talk about graph db a bit but do we have space?}
\label{sec-related-work}
%\walid{The related work section is composed of one big paragraph. You need to break it down to multiple paragraphs, and possibly use guide titles to focus the essence of each paragraph. I introduced some paragraphs as a starter.}
%\Libin{Apri; 11th: added reference to Spruce}
%Graph 
Data systems have been developed for graph data management and analytics~\cite{bytegap-ref,kuzu-ref,orient-db-ref,janus-ref,neo4j-ref,a1-ref,besta2023graph-ref,G-Tran,Weaver,byte-graph-ref,GART-ref,delta-store-htap-ref,vegito-ref,snap-ref,llama-ref,aspen-ref,terrace-ref,
stinger-ref,streaming-sparse-graph-ref,graph1-ref,csr++-ref,risgraph-ref,graphscope-ref,duckpgq-ref,grfusion-ref,dynamic-csr-ref,spruce-ref,gdb-survey,graphflow-ref,lsgraph-ref,columnar-gdbms-ref,gastcoco-ref,cuckoograph-ref,
graphbase-ref,sqlgraph-ref,redisgraph-ref,craygraph-ref,trinity-ref,hypergraph-ref,oracle-graph,tigergraph-ref,sparksee-ref,memgraph-ref,dynamic-graph-streaming-ref}.
%\walid{What is the difference between graph data systems (the previous sentence) and "standalone data systems" in the next sentence?}
%\Libin{Graph data systems include graph databases, graph streaming systems, graph transactional systems, and graph computing engines (read only). I want to emphasize later we work on a specific group of these systems}
Transactional graph systems~\cite{Sortledton,LiveGraph,Teseo} execute read-write transactions and analytics on dynamic graphs concurrently. %on a dynamic graph concurrently. %, e.g.,~\cite{Sortledton,LiveGraph,Teseo}.

LiveGraph~\cite{LiveGraph}, a main-memory graph data system, supports graph analytics and read-write transactions. 
{\color{black}LiveGraph adopts an adjacency list-based edge storage model, and uses a vertex array index storing pointers to vertex versions and edge blocks. LiveGraph supports {\em MVCC}. Read-write transactions acquire exclusive locks on vertices, and create a new version to update a vertex, and append a new update log to update an edge. Readers do not need locks, and directly scan the memory block utilizing transaction timestamps to read the visible versions. LiveGraph stores edge versions of the same source vertex consecutively in the same memory block and enables purely sequential adjacency list scans. LiveGraph's sequential storage of adjacency lists is cache-friendly, reduces random memory access and pointer chasing, and facilitates prefetching~\cite{LiveGraph,Teseo,Sortledton}. It enables efficient graph analytics.}
%LiveGraph's efficient analytics is due to the purely sequential adjacency list scan. It uses $MVCC$ to store edge versions of the same source vertex consecutively in the same memory block. Read-write transactions acquire exclusive locks on vertices to append a new update log to update an edge. Readers do not need locks, and directly scan the memory block utilizing transaction timestamps to read the visible edge versions. LiveGraph's sequential storage of adjacency lists is cache-friendly, reduces random memory access and pointer chasing, and facilitates prefetching~\cite{LiveGraph,Teseo,Sortledton}. 
LiveGraph's edge block lacks indexes, %linear multi-versioned storage lacks indexes within each edge block, 
and scans the whole block for edge lookups and updates (that need to invalidate old edge versions). Also, LiveGraph's vertex-centric locks block concurrent transactions from writing to the same vertex's adjacency list even if they may be updating different edges. {\color{black}Despite competitive graph analytics performance, LiveGraph suffers from low transaction throughput as shown in our experiments in Section~\ref{sec-experiment}. \sys{} uses a similar adjacency list-based {\em MVCC} storage model indexed by vertex ID,  %as in LiveGraph, 
but overcomes these challenges through 1)~A latch-free graph storage (Section~\ref{sec-graph-storage}), 2)~A delta-chains-based edge multi-versioning and index within edge-deltas blocks (Sections~\ref{section:edge-delta-block}, ~\ref{sec-edge-update}-\ref{sec-adj-scan}), 3)~A new high-throughput read-write ACID transaction protocol with delta-chain-based concurrency control (Section~\ref{sec-txn-cc}), 4)~A cooperative safe-timestamp-based garbage collection technique during normal transaction operations (Section~\ref{sec-resource-management}).}
%Teseo~\cite{Teseo} stores dynamic large sparse arrays that represent a graph inside leaf nodes of a ``fat" B$^+$-tree to provide both efficient updates and adjacency list reads.

Teseo~\cite{Teseo} stores graph vertices and their adjacency lists in dynamic large sparse arrays~\cite{sparse-array-ref} inside leaf nodes of a ``fat" B$^+$-tree to provide both efficient updates and adjacency list reads.
Teseo has a clustered sparse index on keys in each sparse array segment to support fast point lookup of vertices and edges, and secondary indexes on vertex locations in the segments to initiate adjacency list scans. Teseo has a hybrid latch {\color{black}per sparse array segment} that combines conventional and optimistic latches to support single-writer multi-reader semantics, and reduces concurrency control overhead in read-intensive workloads. A read-only or read-write transaction can acquire a latch either conventionally or optimistically based on the desired operation and transaction type.
%to access/modify the segment.
{\color{black}Teseo spawns service threads for sparse array and B$^+$-tree structure modifications.}
When a segment in a sparse array is full, Teseo requires a service thread to determine a rebalance window of multiple segments in the sparse array, lock them, and redistribute entries among these segments. 
%\walid{You did not say what is wrong with doing the above?}
A sparse array resize either creates a new sparse array as the leaf or splits the leaf into two new sparse arrays. % (fat tree leaves). 
%\walid{what is wrong with doing that? you did not say.}
These operations require locking multiple memory blocks exclusively. Transactions that do not conflict with normal operations may conflict in locking neighbor segments that may cause threads to block and stall. Also, multiple threads may compete in locking segments for rebalancing. {\color{black}Maintaining service threads also increases computing resource contention. Experiment results in Section~\ref{sec-insertion-exp} show that Teseo suffers the most performance degradation in workloads with hotspots due to contentions during structure modifications. \sys{} overcomes these through 1)~A cooperative design with no service thread except a single commit manager. \sys{'s} structure modification (edge-deltas block consolidation) and garbage collection are handled by worker threads cooperatively during runtime (Sections~\ref{sec-txn-cc} and~\ref{sec-resource-management}). 2)~\sys{'s} %cache-friendly 
state-based block-protection protocol designates one worker thread to clean one memory block to reduce contention. Worker threads do not compete to lock memory blocks (Sections~\ref{sec-block-protection} and~\ref{sec-consolidation}).}
%\walid{You did not explain what "rebalancing" means. I missed that.}
%\Libin{Feb 23rd. Professor, does this part address that "However, when a segment inside a sparse array becomes full, Teseo requires a service thread to determine a rebalance window of multiple segments in the sparse array, lock all of them, and redistribute entries among these segments". It redistributes entries of full segments to other segments to rebalance the loads}
%\Libin{March 13th: is the following sentence still needed?}
%The discrepancy between insert and update performances of Teseo in our experiments further demonstrates that rebalancing and splitting are a major bottleneck in Teseo. 
%Libin Sept 24th: is the below sentence good here?

%Thus, \sys{} designates one worker thread to conduct garbage collection on each adjacency list via \sys{}'s state-based block protection protocol. 

%\Libin{Above addressed professor's comments}
%\walid{Can you briefly state how this issue is avoided in \sys?}
%\Libin{Done!}

Sortledton~\cite{Sortledton} is a graph data structure optimized for the access patterns popular in graph analytics. 
%\walid{By "most relevant", do you mean: popular? or ones that fit real world scenarios? or what?}
%\Libin{They study the access patterns of popular graph analytics algorithms and optimize for those patterns (so I will say both points are valid in your comment)}
%data access patterns in graph computation workloads based on storing adjacency lists in ordered sets. 
It {\color{black}organizes vertices in a 2-level vector index and} stores sorted adjacency lists in %large 
{\color{black} vectors and edge }%memory 
blocks for better read performance at the expense of maintaining the sort during updates.
%to balance 
%the sorting cost in updates and sequential memory access.
%the sorting cost in updates vs. the sequential memory access.
%\walid{is the above minor fix correct? also, you did not state why there is a sorting cost for updates?}
%\Libin{Yes, I meant to say they want to have sorted blocks for better read performance. Therefore every update needs to keep the block sorted}
{\color{black}For hub vertices,} Sortledton has a concurrent unrolled skiplist~\cite{unrolled-skiplist-ref} to store blocks of edges in sorted sets, where each skiplist element is an edge block. {\color{black} Sortledton's transactions support scan and set operations on adjacency lists and concurrent vertex/edge updates. However, Sortledton requires transactions to know in advance their read and write sets, and lock them following a global order beforehand through vertex locking. Transactions need to complete all their reads before doing any writes. \sys{'s} ACID transaction protocol 
%is generalized and 
does not require knowing read sets and write sets beforehand. A \sys{} transaction can execute reads and writes in any order during runtime (Section~\ref{sec-txn-cc}). Moreover, \sys{'s} delta-chains-based concurrency control eliminates vertex-centric locking (Section~\ref{sec:delta-chain-locking}) to provide better concurrency.} %Sortledton has fast scan and set operations on adjacency lists while concurrently updating edges. It uses a read-write latch per vertex, and restricts transactions to know their read and write sets and acquire all latches in advance. Its transactions' reads are to be completed before starting writes.

%In contrast, \sys~ has an adaptive concurrency control protocol with evolving ``locking" granularity based on the workload. 
%On the other hand,
Latch-free indexes and  systems~\cite{BwTree,
%openBwTree,
EPVS,quadboost-ref,lock-free-data-structure-ref, lock-free-slides-ref, Deuteronomy-ref,hekaton-ref,bztree-ref,high-performance-mvcc-ref,tictoc-ref,silo-ref}
use atomic operations, e.g., 
%atomic load, 
$CAS$ and {\em fetch\_add}, to serialize concurrent operations. %\sys~ shares similar ideas with the Bw-Tree~\cite{BwTree, openBwTree}. \Libin{maybe we also share ideas with e.g. hekaton, tictoc in atomic operations. Need to verify here.}
%Libin: Oct 11th: I think the below sentence contains incorrect information. We did not resolve write-write conflict with delta-based multiversionig and lock-free reads. Write-write conflicts are resolved by atomic words (delta-chain locking's lock bits). Therefore I make the following revisions.
%\sys~ uses atomic operations for concurrent reads and writes and resolves write-write conflicts with delta-based multi-versions for lock-free reads. 
\sys~ uses atomic operations and delta-based multi-versions for concurrent lock-free reads and to resolve write-write conflicts.
It has sequential memory blocks for adjacency lists as in~\cite{openBwTree,LiveGraph} to amortize the cost of delta allocations. 
Deltas are stored sequentially to reduce random access. 
%Libin: sure, this is just some lessons we learned from previous works which I thought are "background", but it will show up in later sections as well.
%\sys~ is collaborative as in~\cite{BwTree,EPVS,openBwTree,bztree-ref,quadboost-ref}.
%(Sections~\ref{bwgraph-transaction}, \ref{sec-resource-management}). \\
%Libin: I was thinking about I need to explain a bit more about what collaborative stands for? It will be explained later. So we can remove it right now for space.
%\walid{if you want to remove go ahead, or leave it but add here, more on collaboration will be covered in section....}
%\Libin{I see. Let me think about this a bit more}

%\vspace{-1mm}
\section{Overview of \sys{}}
\begin{figure}[htbp]
    \centering
    \includegraphics[width=1\linewidth]{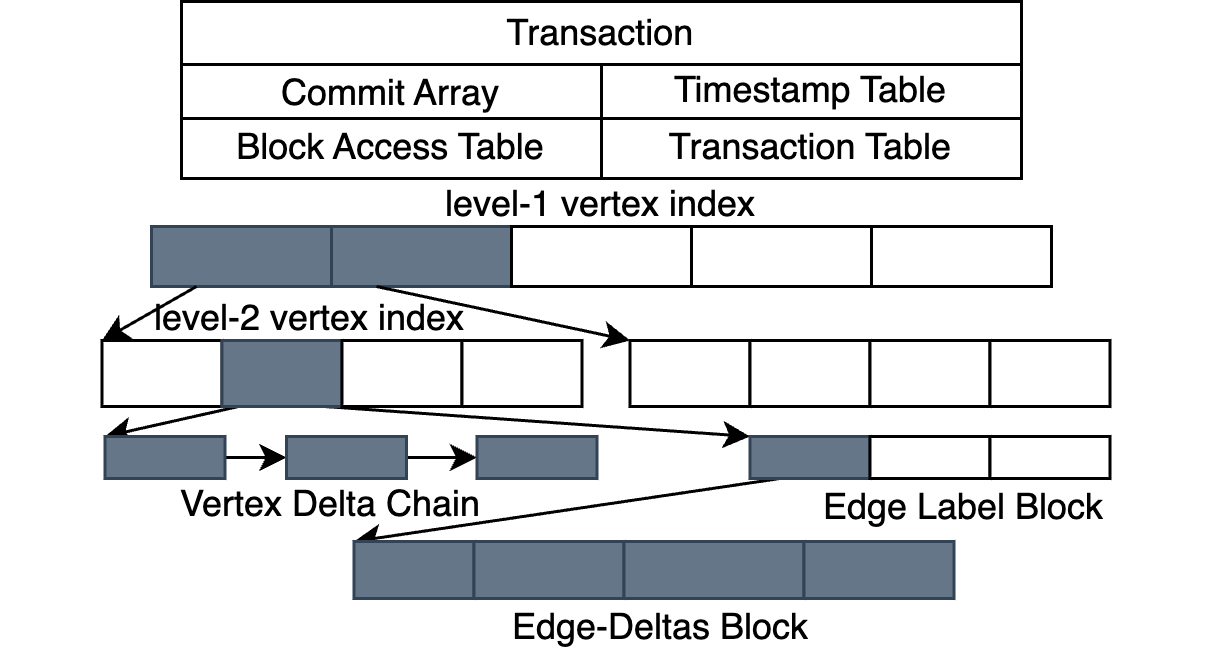}
    \caption{\textcolor{black}{\sys{} Overview: Transactions and graph store}}
    %\vspace{-2mm}
    \label{fig:gtx-overview}
\end{figure}
%Jan. 2nd Libin: add problem definition here?
\label{gtx-overview}
%In this 
%work, we tackle 
%\Libin{March 29th: add reference for property graph and indicate we target the similar type of storage like LiveGraph. Is the "LiveGraph" part necessary? I want to acknowledge that we learn from their storage a bit, but don't want us to be like a mere adaptation.}
%paper, we 
% yeasir: typo? 'dynamically'
%Libin: should be "dynamic labeled property graph", dynamic describes graph not "labeled".
\sys{} manages and queries dynamic labeled property graphs~\cite{pg-ref}, where edges can contain labels, e.g., as in~\cite{LiveGraph}. %Libin: shall we talk more about what LPG is?
Fuchs et al.~\cite{Sortledton} microbenchmark the popular access patterns in graph workloads.  
Major takeaways from their study are: (1)~Storing neighbors of all vertices together (sequential vertex access) is beneficial but is not necessary.
(2)~Sequential access to each vertex's adjacency list enables better performance. 
(3)~A dense vertex identifier domain has 
better performance for graph analytics than a sparse domain. (4)~Some graph algorithms access vertices in random order and require a vertex index.
Given these findings, 
we make several design decisions. % for \sys{}.
\sys{} uses an adjacency-list-like structure, and stores edges 
%(versions) 
of 
a vertex sequentially in memory blocks (Section~\ref{section:edge-delta-block}).
% yr(Referring to sections may help the readers here, e.g., \sys{} uses an adjacency list-like structure, and stores edges (\S{ref:sec-edge-delta})) and later for tx case as well
%Libin: I have added a reference here to the edge deltas block section
\sys{} uses a %n array-based 
vertex index for fast lookup, and an atomic integer on a dense domain to manage vertex IDs. 
For atomicity, consistency, and isolation, graph updates and analytics are executed under read-write or read-only transactions with snapshot isolation~\cite{SI-ref,mvcc-ref}. Transactions support vertex and edge reads, adjacency list scans, and writes (insert, update, and delete). Graph analytics invoke adjacency list scans iteratively via a read-only transaction. \sys{} supports graphs with uniform and power-law edge distribution but is optimized for power-law graphs.
%\lu{Will the optimization for power-law graphs penalize uniform graphs?}
Figure~\ref{fig:gtx-overview} gives an overview of \sys{'s} storage and constructs for transactions. {\color{black} The colored portion in gray corresponds to 
%GTX’s 
\sys{'s} graph storage %in Section~\ref{sec-graph-storage} 
and the uncolored portion 
%describes 
corresponds to \sys{'s} transaction protocol with auxiliary data structures. %in Section~\ref{sec-txn-cc}.
}%\textcolor{red}{The colored portion gives an overview of GTX’s graph storage (to be detailed in Section~\ref{sec-graph-storage}) and the uncolored portion describes transaction protocol with auxiliary data structures (to be detailed in Section~\ref{sec-txn-cc}).} %More detail about each of them are in the following sections.

%\Libin{I simplify the following paragraph due to redundant information}

%\sys{}'s graph storage has a vertex index. Each index entry has 2 pointers that point to the head of a {\em vertex delta chain}, and the first {\em edge label block} of this vertex, respectively. {\em Edge label block} stores pointers to {\em edge-deltas blocks} of different labels. {\em Vertex delta chain} is a linked list of vertex versions while an {\em edge-deltas block} stores all edge-deltas (versions) of the edges of the same source vertex with the same label. Figures~\ref{fig:graph-storage}a and~\ref{fig:graph-storage}b give \sys{}'s  graph structure, a 2-level vertex index-based graph storage, and 1 vertex entry from the 2nd level vector with this vertex's version and edge storage. This design is inspired by LiveGraph~\cite{LiveGraph}. \sys{} manages the following tables that aid its transaction manager and garbage collector: The group commit array, the block access table, the timestamp table, and the distributed transaction table. These global tables allocate one entry per worker thread. Each worker thread has its own local garbage queue and all worker threads share a hybrid block manager. 

%%\vspace{-3mm}
\begin{figure}[htbp]
\includegraphics[width=\linewidth]{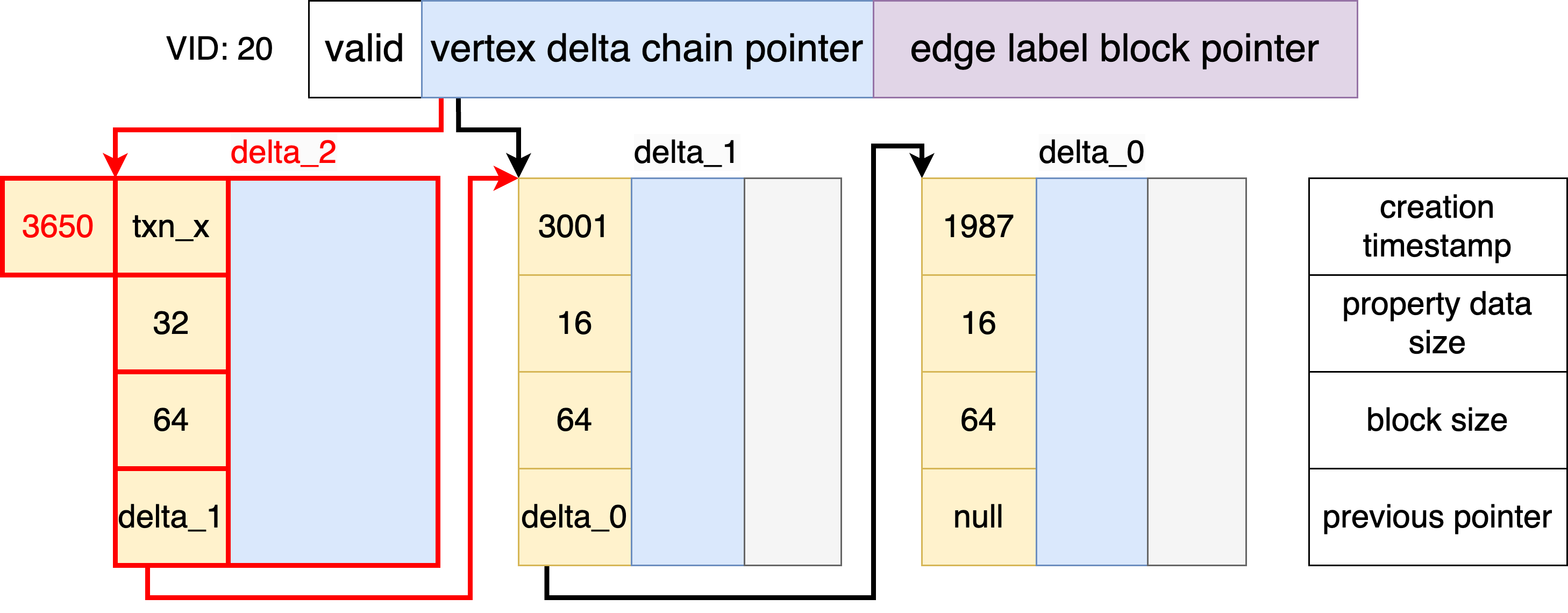}
  \caption{Vertex Delta and Operations}
  %\vspace{-3mm}
  \label{fig:vertex-example}
\end{figure}
\section{\sys{} Storage}
\label{sec-graph-storage}
%\subsection{\textcolor{red}{Delta-based MVCC Storage}}
%\Libin{here I want to briefly discuss our high level idea of delta store/version store and emphasize we choose a matured and well-developed design for this topic.}
%\sys{} adopts Multiversion Concurrency Control ($MVCC$)~\cite{mvcc-ref}. %Like other relational and graph 
%Libin: add a sentence saying we are main memory system?
\sys{} is a main-memory MVCC graph data system. Similar to other $MVCC$-based systems~\cite{high-performance-mvcc-ref,hekaton-ref,cicada-ref,LiveGraph,Teseo,Sortledton}, \sys{} stores vertex and edge updates into deltas, links deltas of the same object, and manages deltas' lifespan and visibility through timestamps. \sys{} uses a 
% yeasir: is "matured" MVCC something different from plain MVCC?
%Libin: no haha, I just meant we used an MVCC that has been adopted in many well-known accomplished systems, hence a matured design.
%matured 
multi-versioning design similar to Hekaton~\cite{hekaton-ref}, Cicada~\cite{cicada-ref}, and LiveGraph~\cite{LiveGraph}. Each of \sys{} delta's visibility and lifespan are managed by creation and invalidation timestamps. A delta's lifespan is the time between its creation and its invalidation timestamps (that is infinity for the latest version delta). A transaction $T$ is assigned a read timestamp ($rts$), and can read all deltas whose lifespan overlaps $T$'s $rts$.
%\lu{Lifespan is a duration and timestamp is a time point. I feel like the word "overlap" is not appropriate.} %Libin: fixed that
%Libin sept 24th: maybe the next sentence is not necessary, we will cover storing ID part in the transaction management section
%When $T$ updates a vertex or an edge, it creates a new delta and stores $T$'s ID in the creation and invalidation timestamps of the new and previous version deltas.
$T$ creates deltas to update vertices and edges.
%(if exists). %Similar to Cicada~\cite{cicada-ref} and LiveGraph~\cite{LiveGraph}, \sys{} supports "transactions reading own writes". Using transaction IDs as timestamps during transaction runtime serves this purpose. 
Deltas of the same vertex or edge are linked %following chronological order 
forming version chains. The beginning of the version chain is the most recent delta. \sys{} enforces that every delta has the same creation timestamp as the invalidation timestamp of the delta right after it. %Such guarantees are $MVCC$ standards and can be seen in other works~\cite{high-performance-mvcc-ref,hekaton-ref,cicada-ref,LiveGraph,Teseo,Sortledton}.

%\vspace{-3mm}
\subsection{Vertex-centric Index}
\label{sec-vertex-store}
%Libin: as discussed, I will work on shrinking this part by emphasizing on what our design motivation and ideas
\sys{} uses a dense vertex domain, and maintains a vertex index. Figure~\ref{fig:gtx-overview}'s colored portion {\color{black}(in gray)} 
%\lu{Should we avoid using colors? Also I feel like figure 1 is a little empty with many empty rectangles in the figure. I think we can show some example values in these rectangles like figure 3 of livegraph paper.}
% yeasir: I think the figure mentioning here is redundant since you already say it in the previous section.
%Libin: I added "colored portion"
illustrates \sys{}'s storage. Sortledton shows the benefits of having vector-based vertex table for graph storage~\cite{Sortledton}. \sys{} designs a similar latch-free two-level vector-based vertex index. 
Level-1 stores pointers to Level-2 in a fixed-size array. Level-2 has %individual 
vertex index vector segments that are of the same size (or growing sizes). 
%Bw-Graph uses atomic 64-bit integers as vertex IDs. 
%WGA: Is the above really important to mention?
%\walid{For the above, shouldn't you reference Figure 2 so the user at least looks at the two levels? Also, in the figure, can first level be on top and the second level be at the bottom? Also, in 2nd level, do you need such a big array? Can you have only two entries, and then add ..., and then show the second level index vector for the second vector ptr of level one? again two entries with ...?}
%\Libin{Yes I will do that!}
A transaction atomically {\em fetch\_adds} a global offset variable, and uses the return value as the ID of a new vertex, say $v$. 
The corresponding Level-2 entry in the vertex index points to $v$ and its edge versions. 
%\Libin{April 14th: I skipped the vertex index figure as its information seems repeating the high level figure}
\iffalse
\begin{figure}[htbp]
\includegraphics[width=\linewidth]{Figures/System_Figures/Vertex Index.png}
  \caption{Vertex Index}
  \label{fig:vertex-index}
\end{figure}
\fi
The initial vertex index has only 1 pointer in Level-1 and allocates 1 Level-2 vector. When the vector runs out of entries, the transaction creating the first vertex that resides in the next vector allocates the next Level-2 vector, and claims its first entry. Other concurrent transactions creating vertices that need to be in the next vector will wait until the corresponding pointer in Level-1 is initialized. 
%This allocation strategy amortizes the cost of allocating new vectors, and has $O(1)$ vertex lookup. 
%Sortledton~\cite{Sortledton} has a similar 2-level structure.
%\walid{if this scheme is borrowed from sortledton, state so in the beginning of it as it is not clear what is from sortledton and what is new here.}. 
%\Libin{I may need to emphasize that part of the structure ideas are inspired from Sortledton's design but our index is not a direct borrow from them}
%Libin: Jan. 8th revision: address first level array runs out of space
%\walid{If the following is your new revision, then please delete the old revision as I am confused, which ones is the next that needs to be in the paper.}
%\Libin{It is the new added paragraph to address an issue raised by the reviewers, but the previous should still stay}
%To accommodate large numbers of vertices, we manage the reference to the 1st level array by an atomic pointer. When the Level-1 array becomes full, the transactions that need to create new vertices will allocate a 2X larger level-1 array, copy all pointers to the new array, and allocates a new vector there. Then, it uses $CAS$ to update the pointer to reference the new array. Failed $CAS$ indicates that another transaction has allocated a new array, and thus other transactions should only reload the pointer. 
%For simplicity, we assume that the Level-1 array is large enough to accommodate all vertices. 
When \sys{} rarely runs out of space to allocate more vertex index entries, it allocates a new Level-1 array with double the size, and copies all vector pointers there similar to dynamic array expansion.
The benefits are two-fold. Breaking the Level-2 vectors into segments managed by pointers reduces memory waste of allocating a single large vector. %and only requires appending a new segment during vector resizing without explicitly copying vector contents. 
Level-2 vector extension avoids explicitly copying/touching vector contents and only requires appending a new segment. %It amortizes vector allocation cost. %, and allows low overhead vertex index resizing. 
Meanwhile, \sys{} still maintains $O(1)$ vertex lookup. 
%Given a vertex ID $v$, its 
$v$'s vertex index entry can be retrieved at $V[i][j]$: $i=v/c$ and $j=v \ mod \ c$, where $c$ is the size of each Level-2 vector segment.

A vertex index entry stores an atomic pointer to the head of the vertex delta-chain. %Figure~\ref{fig:vertex-example} gives an example of vertex 20 (V20)'s delta chain.
%Each vertex index entry has 2 pointers and a valid flag. The first pointer points to the head of a delta chain of vertex versions.
%\wga{Is this repeated from above?}
%\Libin{Yes this sentence is a bit of the repeat for the one above, should we remove one of them?}
%\wga{Yes}
%\Libin{April 21st: I have rephrased it a bit. Do you like it now? I find it difficult to just remove one of them. Also professor, some new thoughts, after reading your previous comment, I feel the next paragraph is repeating the same thing we will discuss in the transaction section. Do you think we can remove it? Or reduce its materials?}
%Each vertex version is stored in a delta that points to the previous version, forming a delta chain. Notice that vertex deltas do not explicitly store invalidation timestamps. This value can be inferred from the vertex delta right after it. 
Figure~\ref{fig:vertex-example} shows an example of Vertex 20 (V20), ignoring the parts in red for now. V20 has 2 versions, $delta\_0$ created and invalidated at timestamps 1987 and 3001 and $delta\_1$ created at timestamp 3001. %: the oldest is created and invalidated at timestamps 1987 and 3001, and the latest is created at timestamp 3001.
Each vertex delta stores how many bytes its vertex property takes, and stores the property data inline. %Note that 
$delta\_0$ does not store its invalidation timestamp and infers it from $delta\_1$'s creation timestamp. Each vertex delta stores a pointer to the previous delta of the same vertex, forming a delta chain.
\subsection{Edge Deltas Storage}
%\Libin{March 14th: got it, but professor in the above texts you used edge delta-block. Do you want me to change them? I will call property data region, and edge property}
\label{section:edge-delta-block}
\iffalse
\begin{figure}
 %\vspace{-1mm} \includegraphics[width=\linewidth]{Figures/edge-deltas block2.png}
  \caption{Edge-Deltas Block Layout}
  \label{fig:delta-block}
\end{figure}
\fi
%\Libin{Feb. 26th, I am thinking about potentially removing this example to save some space.}
\iffalse
\begin{figure}
  \includegraphics[width=\linewidth]{Figures/Transaction Example/combined offset update example.png}
  \caption{Delta Allocation Example}
  \label{fig:delta-allocation}
\end{figure}
\fi
\begin{figure*}[htbp] 
    \centering
     \includegraphics[width=1\textwidth]{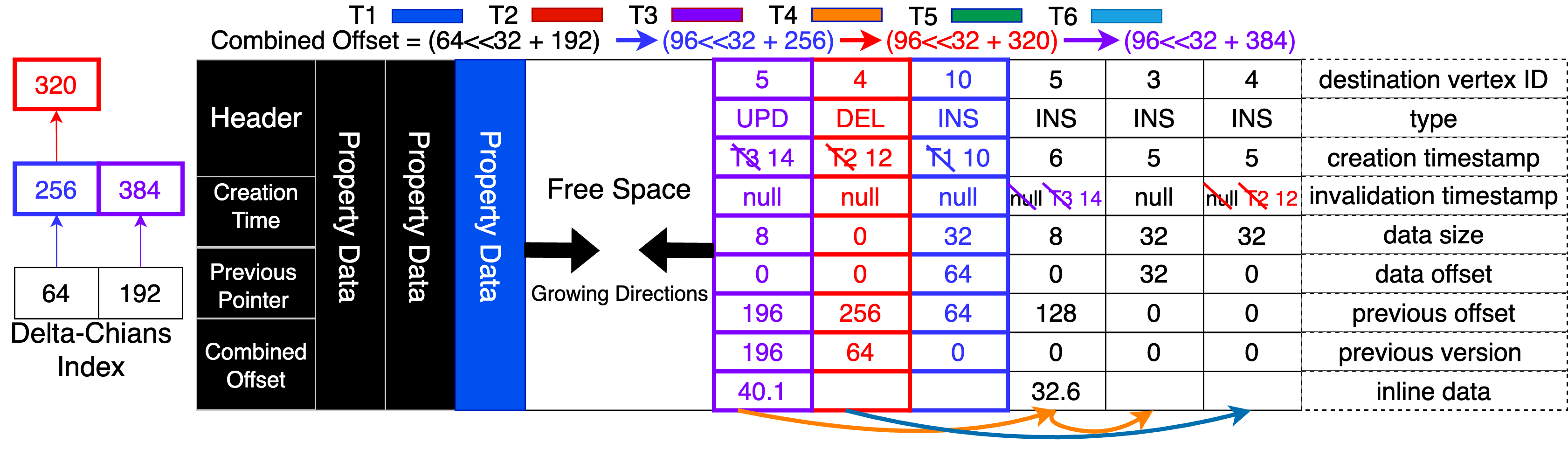}
    %\vspace{-6mm}
    \caption{\textcolor{black}{Edge-deltas Block and Edge Operations}}
    %\vspace{-2mm}
    \label{fig:edge-deltas-block-example}
\end{figure*}
The second pointer in each vertex index entry points to an edge label block.
%to locate the target edge-deltas block of a certain label. 
\sys{} stores all edges of the same source vertex and label in an edge-deltas block. %Due to space limitation, we omit details of how to manage latch-free label blocks. 
It stores one entry per label pointing to an edge-deltas block
%More fields in the label block and label entries are added to support latch-free operations, delta-chains index and edge-deltas block state protection. 
and its associated delta-chains index, version number, and state variable (Detailed in Sections~\ref{sec:txn-operations}, \ref{sec-txn-cc}). \textcolor{black}{It manages all label entries in an array. We use 2 atomic variables per edge label block to manage and access the block in a latch-free manner. The first atomic offset variable controls which entries are valid in the array, and the other atomic pointer points to the next block to allocate more labels (initially null). A transaction allocates a new label entry by incrementing the atomic offset, and uses the corresponding array entry to store the new label. If the array is full, it will allocate the next block and update the pointer. A transaction looking up an label entry simply scans the array. To accommodate concurrent updates, it reads the offset before and after the scan, and will reinitiate the scan if the offset has been updated. }
The experiment datasets and our latter discussion assume the graph contains only 1 label and omits showing labels %them 
for brevity.

Figure~\ref{fig:edge-deltas-block-example}'s contents in black illustrate V20's edge-deltas block and delta-chains index. %Again we only focus on the parts in black for now. 
%\sys{}'s edge-deltas block has 4 sections similar to LiveGraph~\cite{LiveGraph}. 
\sys{}'s edge-deltas block has similar storage layout to that of LiveGraph~\cite{LiveGraph} and improves it with atomic memory allocation, edge delta-chains, and hybrid property data storage.
At the lowest address, the header stores the block's metadata, e.g., creation time.
Then, a property data storage section stores each edge delta's variable-sized property data. 
%that grows after the header from lower to upper address. 
Fixed-sized edge deltas are stored at %grow from 
the opposite end of the block. 
Edge deltas are logically linked in different edge delta-chains.
Each block has a delta-chains index whose $i^{th}$ entry stores the offset of the $i^{th}$ edge delta-chain's head.
The {\em combined\_offset} %in the header 
combines the offsets of the latest edge delta and property data in a 64-bit integer 
%\Libin{question here about reference, shall we reference the other works like silo and tictoc that also combine multiple objects in a single atomic word and manage it atomically?} yes; Libin: I added it.
%Since 64-bit is the largest word 
that can be atomically updated by the hardware without a latch. It enables atomically allocating memory for both the edge delta and its property in a single operation. Its higher 32 bits store the offset of the property data region and the lower 32 bits store the offset of the edge deltas region, extending from single direction space preallocation in Open Bw-tree~\cite{openBwTree}, and latch-based allocation in LiveGraph~\cite{LiveGraph}. 
Silo~\cite{silo-ref} and Tictoc~\cite{tictoc-ref} also store multiple objects in a single atomic word to enable atomic multi-reads and multi-writes. 

Each edge delta takes 1 cache line (64 bytes) to avoid false sharing among concurrent writers. 
%\walid{The combined-offset calculation is detailed, and it is not clear what you are trying to show at the higher level. You are trying to achieve what? State that first, and then we can decide whether these details are worth it or not.}
%\Libin{Fixed, I have stated the motivation and reasons of using this combined offset}
%\Libin{After Jan. 17th meeting: professor wants me to think about simplifying this part. whether we only need to present the highlevel ideas here.}
It stores a write operation %the operation the edge delta stands for, 
and the creation and invalidation timestamps for multi-versioning. An edge delta's {\em previous\_offset} %is the key of the sequential edge delta chains storage, and 
points to the previous edge delta on the delta chain. At an edge-deltas block's creation time, \sys{} determines its edge delta-chain number, where transactions append edge deltas 
to different delta chains using a hash function. 
%Libin: emphasize a bit on similarity and difference among systems
%\textcolor{red}{
Hekaton~\cite{hekaton-ref,high-performance-mvcc-ref} %and its predecessor~\cite{high-performance-mvcc-ref} 
also maintains linked lists of all versions in the same hash bucket (may be different objects). %In the most ideal scenario that each key lies in a separate hash bucket. Hekaton links versions of the same key together, and its hash index becomes indexing version chains by keys. 
%While sharing the similarity in delta chains, 
\sys{'s} novelty lies in the combination of pointer-based delta linking and sequential storage preallocation. Its delta linking serves as an index for efficient single edge delta lookup %without sacrificing the sequential storage. 
and all edge deltas of the same vertex are still stored sequentially in one memory block to preserve localities in adjacency list scans. Such a locality-preserving sequential storage proves to be beneficial for the graph analytics~\cite{tree-store-graph-ref,LiveGraph,Teseo,Sortledton,spruce-ref}. %that cannot be found in Hekaton. %Robust graph analytics performance in graph systems using sequential neighborhood storage proves to be beneficial~\cite{LiveGraph,Teseo,Sortledton,spruce-ref}.
Finally, \sys{} delta chain storage and delta-chains index are tightly coupled with concurrency control (Sction~\ref{sec:delta-chain-locking}). 
%2024 Sept 10th: Libin: I added the below sentences in red to emphasize the conversation a bit more on delta chain locking
%\textcolor{red}{Delta chains are locked and ingest new edge deltas in parallel to break down vertex-centric locking. Each edge deltas block also adjusts its delta chain number at runtime to modify its concurrency level to adapt to the workloads.}
Each edge delta also 
stores the offset of the same edge's previous edge delta to support efficient visible version lookup. Given a target edge $e$, a reader transaction locates $e$'s edge delta-chain, 
and uses
edge deltas' {\em previous\_offset} and  {\em previous\_version\_offset} to locate $e$'s visible version. Edge delta's data size and data offset locate its edge property data. We extend Cicada's best-effort inlining~\cite{cicada-ref} and take advantage of ``wasted" space in each cache line-aligned edge delta to design a {\em hybrid} storage scheme for its variable-sized property. If the property size is small, e.g., $\leq 16$ bytes, it is stored in the cache line-aligned edge delta to save space and reduce random memory access, else it is stored in the data region. %through storing a data offset and data size. 
In contrast to Cicada's inlining that only applies to one version and has management overheads, every \sys{} edge delta can store property data inline to save space and improve cache performance. Because each edge delta's inline data region is allocated regardless of the property size for cache-line alignment, there is no extra overhead for \sys's hybrid storage scheme.
% yr: The explanation is a bit difficult to follow with the image. E.g., the labeling of edge (20,4), (20,3), (20,5) is not clear. which one are edge deltas? Also, do you think putting this at the very beginning might help? Also, having multiple paragraphs for this section will help. It's pretty dense now. 
%Libin: when you msay "labeling of (20,4), (20,3), (20,5)", do you want me to point out they are represented by deltas? I have revised the writing a bit. Please let me know if this is what you meant. I have removed a few words and break the main part into a few paragraphs. I'm not sure about the location of the figure yet. I feel it is close to the text enough, and I feel putting it in the previous page will mess up the organization of the other 2 figures in that page. Do you have an idea how to make it better?
% yr: I was thinking of labelling the last 3 columns, say they are edges and the prev 2 are deltas. But, I don't think this is an issue. I meant the location of the example writing. It's at the very end. I will rather prefer it at the very beginning maybe the first or seond paragraph. 
%Libin: thank you and now I see your point. Do you mean to say labeling/saying the columns on the right as edge deltas? I don't know if there is an easy way. I see your point about the example. I'm wondering if it will be an issue becuase we haven't introduced what edge deltas and the block structure are. Will the readers find these confusing? 

In the example, V20 has 3 edges: $e(20,4)$, $e(20,3)$, and $e(20,5)$ 
($insert$ edge deltas)
and maintains 2 edge delta-chains. 
Its delta region takes 192 bytes and data region takes 64 bytes ({\em combined\_offset}). $e(20,4)$ and $e(20,3)$ have 32 bytes of property data each stored in the data region (Offsets 0 and 32). $e(20,5)$ stores its 8-byte weight inline in the edge delta. $delta\_chain\_0$ contains 1 edge delta $d(4, INS)$ at Offset 64, and $delta\_chain\_1$ contains 2 edge deltas $d(3, INS)$ and $d(5, INS)$ starting at Offset 192.
The middle section is free space.
\section{Graph Transaction Operations}
\label{sec:txn-operations}
%\Libin{I have added 1 sentence highlevel summary of each graph operation at the beginning.}
\sys{} transactions support vertex and edge insert, delete, and update, single vertex/edge lookup, and adjacency list scan.
In this section, we persent transaction operations, and defer transaction management (e.g., concurrency control, commit) to Section~\ref{sec-txn-cc}. Let $cts$, $its$, $rts$, and $wts$ be the creation, invalidation, read, and write/commit timestamps, respectively.
At creation time, each transaction gets an $rts$ %($rts$) %from the $global\_read\_ts$ 
and can see all deltas with overlapping lifespans. 
For writes, transactions create deltas with transaction IDs as timestamps, and update them with $wts$ at commit time.
%\vspace{-3.5mm}
\subsection{Vertex Read and Update}
\label{sec-vertex-operations}
\sys{} transactions update vertices through creating vertex deltas, and read them by finding/reading visible deltas using atomic load and $CAS$.
%Refer to the red parts in the same example of V20 in Figure~\ref{fig:vertex-example}.
Refer to the same example of V20 in Figure~\ref{fig:vertex-example} including the red parts.
%Let's ignore the parts in red first. 
%This example shows v
%V20 has 2 versions, $delta\_0$ and $delta\_1$: the oldest is created and invalidated at Timestamps 1987 and 3001, and the latest is created at Timestamp 3001. Each vertex delta stores how many bytes it uses as vertex property, and stores the property data in the delta. Also, each vertex delta stores a pointer to the previous delta of the same vertex, forming a delta chain. 
Assume that Transaction $t$ with $rts=2000$ looks up V20. After fetching V20's delta chain pointer from the vertex index, $t$ checks the delta chain head $delta\_1$, and finds its $cts$ is larger than $t$'s $rts$. Thus, $t$ uses the previous pointer to locate $delta\_0$ with $cts=1987$. $t$ determines $delta\_0$ is visible, and reads it. If a newer transaction $t2$ with $rts=4000$ also reads V20, $t2$ will read $delta\_1$. %the vertex delta with $cts=3001$. 
Assume that a read-write transaction $txn\_x$ updates V20. Similarly, it loads the delta chain head as in the read operation. $txn\_x$ checks $delta\_1$'s $cts$, and makes sure it is not a transaction ID, nor larger than $txn\_x$'s $rts$. Else, $txn\_x$ immediately aborts due to a write-write conflict. $txn\_x$ creates a new vertex delta $delta\_2$ with the updated property, and stores $txn\_x$ as its $cts$. $delta\_2$ stores a pointer to $delta\_1$. Then, $txn\_x$ invokes $CAS$ to update the vertex delta-chain pointer to point to $delta\_2$. If $CAS$ fails, $txn\_x$ aborts due to write-write conflict. Else,  $txn\_x$ commits at $wts=3650$, and updates $delta\_2$'s $cts$ to its $wts$. This applies to vertex insert and property update. Deletes are covered in Section~\ref{sec-vertex-deletion}.%This concludes the vertex update and read operations. 

While using a pointer-based delta chain and $CAS$ is less efficient due to pointer chasing, random memory access and competing $CAS$~\cite{openBwTree}, vertex updates are less frequent than edge updates~\cite{LiveGraph}, hence less conflicts from concurrent updates. Also, most transactions need access to only the latest version of a vertex, and will access it with a single pointer chase. Thus, the negative effects of this design are minimal. %In contrast, vertex delta chains is beneficial.
On the other hand, detecting write-write conflicts to the same vertex becomes simple. A failed $CAS$ indicates a concurrent transaction has updated the vertex delta chain.
%Due to the complex nature of vertex deletes (ensuring no dangling edges and cascading deletes), we come back to it after we introduce edge operations.
\iffalse
\begin{figure*}[htbp] 
    \centering
     \includegraphics[width=1\textwidth]{Figures/example_based_writing/edge_operations_example.png}
    \caption{Edge Operations}
    \label{fig:edge-operations}
\end{figure*}
\fi

%\vspace{-3mm}
%2024 Sept 11th: Libin: I have moved delta chain locking to a separated section at 6.1 It does look better but my concern is whether we will be in trouble to talk about "locking the delta chain" without talking about how to lock it.
\subsection{Edge Update}
\label{sec-edge-update}
%Now we exhibit how \sys{} transactions execute edge update operations. 
\sys{} transactions update edges by
%locking edge delta chains and 
creating edge deltas (insert, update, and delete) in edge-deltas blocks. 
%\textcolor{red}{\sys{} proposes adaptive delta chain locking: write-write conflicts are detected at delta chain granularity, and edge-deltas blocks determine edge chain number according to the workload history.} \sys{} supports directed and undirected graphs. 
%Based on different requirements, 
\sys{} supports directed graphs %either by only storing edges (deltas) in source vertex's edge-deltas block, or use 
using 2 labels and 2 edge-deltas blocks per vertex for fanin and fanout edges. %Libin: I will reference all other systems doing the same thing here
Undirected graphs store each edge twice in both vertices' edge-deltas blocks as in~\cite{Teseo, Sortledton}. 
%\sys{} transactions support updating multiple edges atomically, so supporting undirected graph is trivial. 
%We focus on undirected graphs while in this example we only show operations on one vertex's adjacency list for simplicity (the reverse part is identical).

%Figure~\ref{fig:edge-operations} gives an example of V20's adjacency list. We omit the labels for simplicity. Assume the parts in black refer to the original adjacency list before any updates. V20 has 3 edges: $e(20,4)$, $e(20,3)$, and $e(20,5)$ and maintains 2 edge delta chains. Edges $e(20,4)$ and $e(20,3)$ have 32 bytes of property data each stored in the data region. $e(20,5)$ has an 8-bytes weight, and stores it inline in the edge delta. $delta\_chain\_0$ contains an edge delta $(4, INS)$ at Offset 64, and $delta\_chain\_1$ contains two edge deltas $(3, INS)$ and $(5, INS)$ starting at Offset 192.
Refer to the example of V20's edge-deltas block in Figure~\ref{fig:edge-deltas-block-example}. Colors refer to transactions' updates, e.g., Transaction $T1$'s updates are in dark blue.
Assume that $T1$ with $rts=9$ updates $e(20,10)$ with a new 32-bytes property. First, it  calculates that $e(20,10)$ belongs to $delta\_chain\_0$ by $10 \ mod \ 2 = 0$, where  the number of delta chains in this block is 2. $T1$ reads the delta-chains index, finds the offset 64, and checks the edge delta at this location. $T1$ sees $d(4, INS)$ has a smaller $cts$ than its $rts$, so it decides to lock $delta\_chain\_0$ (covered in Section~\ref{sec:delta-chain-locking}). 
After $T1$ successfully locks $delta\_chain\_0$, it uses each edge delta's {\em previous\_offset} to look for the previous version of $e(20,10)$, that in this example 
%doesn't 
does not
exist. A lookup of the previous version is necessary to avoid creating duplicate edges and link versions together. Delta-chains index enables efficient lookup of edges.
%Thus, $T1$ determines it will create an edge insertion delta.%its update is indeed an insertion. 
%\walid{What do you mean by the previous sentence? also, the next sentences below are unclear. Need to be rewritten for clarity. If you mean adaptivity based on size, you need to be clear.}
%Sept 10th Libin: I rephrased it a bit. We more of less just want to make sure we tell the users if the edge already exists or not for each update. I think we can also just omit this sentence. For the below sentence, I was referring to hybrid storage of edge deltas.
Since the new edge version has a 32-bytes property, $T1$ needs to store the property in the data region. Then, it invokes the delta allocation protocol (32 bytes for data and 64 bytes for delta regions) to update the {\em combined\_offset}, and creates the corresponding edge delta with $cts=T1$, $type=INS$, {\em previous\_offset} pointing to $d(4,INS)$, no {\em previous\_version\_offset}, data size equaling 32, and data offset equaling 64. When $T1$ is committing, it updates the delta-chains index entry to point to $d(10,INS)$, releases the delta-chain lock, gets its $wts=10$, and updates $d(10,INS)$'s $cts=10$. %More details on transaction commit are in the next section~\ref{sec-hybrid-commit}.

%Sept 12 2024: Libin: I significantly shrink the next paragraph when introducing T2 and T3. We spent enough efforts on details of T1. We just discuss the differences of T2 and T3 here.
The example also shows 2 other transactions $T2$ and $T3$ deleting $e(20,4)$ and updating $e(20,5)$ with a weight 40.1. $T2$ follows similar procedures as $T1$ with a few differences. $T2$ deletes an existing edge, so it fetches $d(4,INS)$ and puts transaction ID $T2$ in $d(4,INS)$'s $its$. $T2$ also writes $d(4,INS)$'s offset 64 in $d(4,DEL)$'s {\em previous\_version\_offset} to link versions. During commit, $t2$ writes its $wts=12$ to $d(4,DEL)$'s $cts$ and $d(4,INS)$'s $its$. $T3$ \textbf{updates} $e(20,5)$ with a new 8-bytes weight and works similarly as $T2$. The difference is that the new weight is stored inline in $d(5,UPD)$. $T1$ stores it in the data region and $T2$'s $d(4,DEL)$ has no property data.
\subsection{Edge Lookup}
\sys{} supports single edge lookup given $(src, dst, label)$. 
The $src$ and $label$ locate the edge-deltas block and its delta-chains index. Then, each reader transaction $t$ calculates the edge delta-chain from $dst$, and reads its deltas. After finding a delta with a matching $dst$, $t$ compares its $rts$ with the delta's lifespan, and uses the {\em previous\_version\_offset} to locate the visible version if needed.

Refer to the same example in Figure~\ref{fig:edge-deltas-block-example}. %Again we omit the label part for simplicity. 
The colors of arrows represent the search path of each transaction. Transaction $T4$ with $rts=15$ looks up $e(20,3)$. After locating the edge-deltas block, % and its delta-chains index, 
$T4$ calculates $e(20,3)$ belongs to $delta\_chain\_1$ by $3\ mod \ 2 = 1$. $T4$ reads the offset (384) from the index, and finds $d(5, UPD)$, the head of $delta\_chain\_1$. %Because $T2$ has no interest in $(20,5)$, 
Then, $T4$ reads $d(5, UPD)$'s {\em previous\_offset} to search the delta chain. After reading $d(5,INS)$’s {\em previous\_offset}, $T4$ finds $d(3,INS)$, and checks its $cts$ and $its$. $T4$ determines $d(3,INS)$ is visible, and uses its data offset 32, and the data size 32 bytes to read the edge property accordingly. Transactions $T5$ with $rts=15$ and $T6$ with $rts=10$ both look up $e(20,4)$. %Both transactions locate the edge-deltas block, read the delta-chains index, and loads $delta\_chain\_0$ at $(4,DEL)$. 
After reading $d(4,DEL)$ of $delta\_chain\_0$, $T5$ determines that $d(4,DEL)$ is visible, and reports $e(20,4)$ being deleted. $T6$ sees $d(4,DEL)$'s $cts=12$, larger than its $rts=10$. Therefore, it uses $d(4,DEL)$'s {\em previous\_version\_offset} 64 to find $d(4,INS)$, verifies its $rts$ overlaps with $d(4,INS)$'s lifespan, and reads the 32 bytes property data of $e(20,4)$.
%\vspace{-2mm}
\subsection{Adjacency List Scan}
\label{sec-adj-scan}
\sys{} supports sequential adjacency list scan given $(src, label)$. After locating the edge-deltas block, a transaction loads the {\em combined\_offset} to determine where to start the scan, and scans every edge delta in the block. It compares its $rts$ with each delta's lifespan, and returns visible edge deltas. We use an iterator model for adjacency list scans, and graph analytics directly accessing each block. In Figure~\ref{fig:edge-deltas-block-example}, Transaction $T7$ with $rts=20$ scans V20's adjacency list. It loads the {\em combined\_offset}, determines the edge deltas start at Offset 384, and reads every edge delta from there. It reports V20 has edges $e(20,5)$ with Weight 40.1, $e(20,10)$ and $e(20,3)$.
%\vspace{-2mm}
\subsection{Vertex Delete}
\label{sec-vertex-deletion}
Vertex deletion is an expensive but necessary operation in graph systems. 
Recent graph database benchmarks LDBC-SNB-Interactive-v2~\cite{ldbc-snb-int-v2-ref,ldbc-snb-int-v2-code}
and LDBC-SNB-BI~\cite{ldbc-snb-bi-ref,ldbc-snb-bi-code-ref} 
include deep cascading deletion operations. 
%Specifically, they require deep cascading deletion~\cite{ldbc-snb-int-v2-ref,ldbc-snb-bi-ref}. 
Deleting a vertex requires removing all its adjacent edges atomically. It preserves graph consistency by avoiding dangling edges.  %\sys{} implements vertex deletion to satisfy privacy laws like European Union’s General Data Protection Regulation (GDPR) that require timely deletion of user data~\cite{ldbc-snb-bi-ref,ldbc-snb-int-v2-ref,gdpr-impact-ref}. On the other side, it preserves graph consistency by avoiding dangling edges. 
%Because vertex deletion needs to happen concurrently with other graph updates and reads, we only briefly describe its procedure but leaves the concurrency part in the next section~\ref{sec-txn-cc}.
%Note that edge deletion has already been covered in the edge update section~\ref{sec-edge-update}. 
\textcolor{black}{Therefore, \sys{}'s vertex deletion contains three steps: (1)~Vertex Locking, (2)~Vertex Delete, and (3)~Edge Delete. Vertex Locking puts the vertex in an exclusive state, and thus preventing other transactions from reading or writing the vertex or its edges. Vertex Delete deletes the vertex, and Edge Delete deletes all its edges.}
Here, we 
%give an overview of deleting vertex
illustrate how to delete Vertex
V20. To delete V20, Transaction $td$ puts  V20 in an exclusive delete state, \textcolor{black}{and installs a special vertex-delete delta. At this moment, V20 and edges in its edge-deltas blocks are considered as being deleted. Then, $td$ scans V20's adjacency list (edge-deltas block) to identify its edges and} deletes these ``reverse" edges in V20's neighbors' adjacency lists using edge delete operations (Section~\ref{sec-edge-update}). \textcolor{black}{$td$ also adds V20 into its thread-local recycled vertex list. After the vertex's deltas and edge-deltas blocks have been garbage collected (detailed in Section~\ref{sec-resource-management}), V20 can be reallocated to accommodate future workloads.} \sys{} currently only allows one vertex delete  at a time, and we find this enough for vertex delete requirements, and it simplifies concurrency control. LDBC-SNB benchmarks have less than $1\%$  writes as deletes, and vertex deletes are a minority among them~\cite{ldbc-snb-bi-ref,ldbc-snb-int-v2-ref,snb-dataset-ref}. %Libin: the paper writing doesn't speicifically say the percent of deletes as edge vs vertex updates. So I downloaded the source dataset, and examine the delete data stream. Edge deletes make up 154MB and vertex deletes make up 45.8MB, so I make the conclusion here. I did reference both the paper and datasets.
More  on vertex delete, including exclusive delete state, is presented in Section~\ref{sec-state-aware-operations}. 
%%\vspace{-1mm}
\section{Transaction Management}
\label{sec-txn-cc}
%\sys{} allocates a user-defined fixed number of worker threads that execute transactions though the worker thread number can change upon user requests during runtime.
\sys{} allocates a set of worker threads that execute transactions.
Each thread executes 
its own
transactions while being cooperative, i.e., performing work on behalf of other transactions.
%of the other threads.
%threads' transactions. 
For analytical workloads, \sys{} uses OpenMP~\cite{openmp-ref} threads to collectively execute a single read-only transaction. % managed by an internal worker thread.
\sys{} supports ACID transactions. \sys{} transactions can execute multiple read and write operations atomically and ensure no dangling or duplicate edges.
\sys{} supports Snapshot Isolation (SI) transactions~\cite{SI-ref} based on $MVCC$~\cite{mvcc-ref}, the only supported isolation level of LiveGraph~\cite{LiveGraph} and Teseo~\cite{Teseo}, 
and stronger than the state-of-the-art graph DBMS Neo4j's default Read Committed~\cite{neo4j-isolation-ref}. We find SI sufficient for graph workload. 
State-of-the-art graph benchmarks LDBC-SNB-Interactive-v1~\cite{ldbc-snb-int-v1-ref} and v2~\cite{ldbc-snb-int-v2-ref} require read committed and SI, respectively. 
%We also implement {\bf serializable} transactions using read set validation~\cite{silo-ref,hekaton-ref,cicada-ref}. We omit the details due to space, but they can be found in \sys's anonymous technical report~\cite{anonymous-gtx-tech-report}. In this section, we focus on SI.
%The original LDBC-SNB-Interactive~\cite{ldbc-snb-int-v1-ref} only requires read committed, and the more recent LDBC-SNB-Interactive-v2~\cite{ldbc-snb-int-v2-ref} extends the requirements to SI. 
\sys{} manages two variables for SI and MVCC: Global read epoch($gre$) and Global write epoch($gwe$) as in LiveGraph~\cite{LiveGraph}. %LiveGraph uses similar variables for SI and MVCC~\cite{LiveGraph}. 
%Libin: I still feel the below sentence is a bit unnecessary.
Silo~\cite{silo-ref} designs an epoch-based transaction protocol but \sys{} is only  similar in resource management.
A \sys{} transaction acquires its $rts$ from $gre$ when created, and gets assigned a $wts$ from $gwe$ at commit time. 
Under SI, \sys{} transactions have no read-write conflicts, and only need to detect write-write conflicts. % covered in sections~\ref{sec-vertex-operations},~\ref{sec-edge-update} using atomic $CAS$ and lock bits. 
%Thus in this section, we focus on vertex and adjacency list level concurrency control.  %concurrency issue raised by vertex deletion and block consolidation. 
%We have covered using 
As in Section~\ref{sec-vertex-operations}, %\walid{give section number} 
we use $CAS$ to detect vertex update conflicts. 
%Later, 
In Sections~\ref{sec:delta-chain-locking},~\ref{sec-block-protection}, %\walid{give section number}, 
we discuss concurrency control for edge delta-chains and edge-deltas blocks.
%\sys{} transactions can read own writes, and only update the latest versions.
Every transaction can see all deltas created,  but not invalidated by itself by storing and checking transaction IDs in delta timestamps. %(by checking creation and invalidation timestamps). 
%Storing transaction IDs as vertex and edge deltas' timestamps makes it easy to detect transactions' own writes. 
%To only update latest versions, a 
A \sys{} transaction $t$ only updates the latest versions: 
$t$
%needs to compare 
compares 
the latest delta's $cts$ of the object $t$ intends to update. $t$ can update only if the latest delta is not an in-progress delta of another transaction and the delta's lifespan overlaps   $t.rts$.
%\walid{do not use "its" lifespan and with "its" as it is confusing above. Say what "it" is. Also, what is "covers"? unclear. please fix!}%Libin: when I say cover, I meant for creation_ts < rts < invalidation_ts. I have changed the term to overlap and change "its rts" to "the rts". I also used the phrase "whose lifespan covers t’s rts." at the beginning of section 4 (introducing multi-versioning). Lu thinks the rts is a point and the lifespan is an interval so "overlap" is not an appropriate word. Should I add a bit more discussion there on what "covers" mean?
\sys{} supports redo logging and checkpoint for durability. 
%In the following subsections, we cover different \sys{} transaction management (especially concurrency control) protocols to ensure high throughput concurrent transactions. 
%Sept 11th: a separate section for delta chain locking
%\vspace{-2mm}
\subsection{Adaptive Delta-Chain Locking}
\label{sec:delta-chain-locking}
Section~\ref{sec-edge-update} shows that a transaction $t$ needs to lock edge delta-chains before installing edge deltas. Locking delta chains enables concurrent updates to different edges in the same edge-deltas block and detect write-write conflicts: Updates to the same edges. Delta chain granularity locking is a middle ground between vertex and edge locks; providing better concurrency than vertex locks and reducing overheads by managing a lock per edge. \sys{} does not use a high-overhead lock manager but a locking bit in the delta-chains index offsets as in Silo and TicToc~\cite{silo-ref,tictoc-ref}'s version timestamps. Each delta-chains index entry uses its most significant bit as the locking bit. To lock a delta chain, $t$ checks if the locking bit is set, and uses a $CAS$ to set the bit atomically. Using locking bits in index offsets enables updating both the locking bit and the delta chain head atomically during commit as in Silo~\cite{silo-ref} and TicToc~\cite{tictoc-ref} that release locks while installing new timestamps. If the bit has already been set by other transactions, or the $CAS$ fails, $t$ detects a write-write conflict and aborts. Section~\ref{sec-consolidation} shows delta chain locking's adaptivity to edge workload history.
%\vspace{-3mm}
\begin{figure*}[htbp] 
    \centering
    \includegraphics[width=1\textwidth]{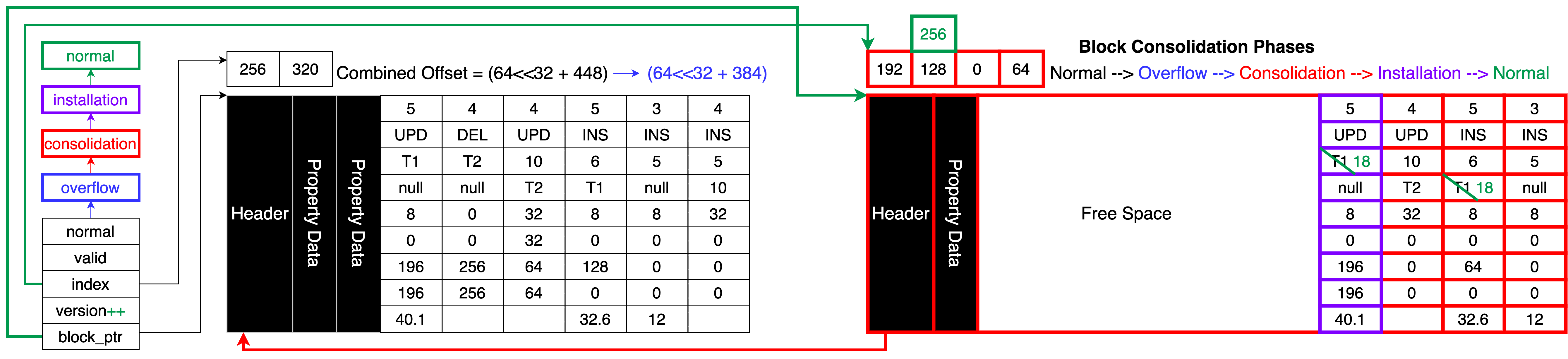}
    \caption{Consolidation Example}
    \label{fig:cons-example}
\end{figure*}
\begin{figure}[htbp]
%  %\vspace{-3mm}
    \includegraphics[width=1\linewidth]{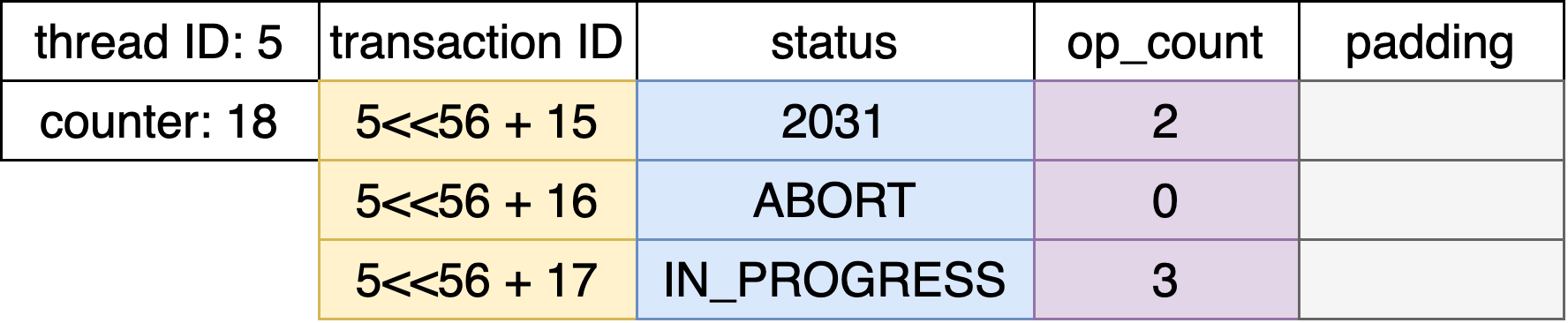}
  \caption{Thread 5's Transaction Table Partition }
  %\vspace{-3mm}
  \label{fig:txn-table}
\end{figure}
%%\vspace{-1mm}
\subsection{Block Protection Protocol}
\label{sec-block-protection}
%To handle certain tasks, 
\sys{} supports different access permissions on edge-deltas blocks using a block state-based protection protocol inspired by~\cite{EPVS}. 
%\sys{} supports different access permissions on edge-deltas blocks. Inspired by~\cite{EPVS}, \sys{} has a block state-based protection protocol that does not use shared-exclusive locks for this task. %, and enables customized concurrency level. 
\sys{} manages a global data structure $block\_access\_table$ with one entry per worker thread. When a worker thread accesses an edge-deltas block, it calculates the block ID through the block's source vertex ID and label, and stores it in its table entry. Each table entry is padded to take one cache line to avoid false sharing. %, and shows what block this worker thread is accessing. 
\sys{} associates every edge-deltas block with an atomic state variable. \sys{} defines 5 states: $normal$, {\em overflow}, $consolidation$, $installation$, and $delete$. $normal$ makes up the majority of an edge-deltas block lifetime where both reads and writes  are allowed. All examples in Figure~\ref{fig:edge-deltas-block-example} assume blocks in State $normal$. States {\em overflow}, $consolidation$, and $installation$ are  for edge-deltas block consolidation (Section~\ref{sec-consolidation}). {\em overflow} and $installation$ provide mutual exclusion on an edge-deltas block, and $consolidation$ makes the block read-only. $delete$ is used for vertex delete, and prevents other transactions from accessing a block. %in Section~\ref{sec-state-aware-operations}. %which we will discuss in Section~\ref{sec-state-aware-operations}. %We will revisit the concurrent vertex deletion in section~\ref{sec-state-aware-operations}.
The Block Protection protocol has 2 operations: {\em register\_access} and {\em change\_state}. When  Transaction $t$ accesses Edge-deltas Block $b$, $t$ calls {\em register\_access(b)} to examine $b$'s state. If $t$'s operation is compatible with $b$'s state, $t$ stores $b$'s ID in $t$'s $block\_access\_table$ entry. Then, $t$ rechecks $b$'s state. If it has not changed, $t$ can 
%execute its operation. 
proceed.
If $b$'s state is incompatible with $t$'s operation, $t$ unregisters its block access by storing 0 in the table, and retries later. Upon finishing, $t$ unregisters its block access. A worker thread $th$ needing to change $b$'s state updates the state to the new value. If the new state is exclusive, %does not require mutual exclusion, $th$ proceeds. Else, 
$th$ scans the {\em block\_access\_table}. If $th$ finds no threads accessing $b$, $th$ executes the operations in the new state. Else, $th$ continues scanning the $block\_access\_table$ until $b$ is ``safe". %The correctness of this protocol is guaranteed by checking the state before and after registers the access. 

Similar functionalities can be achieved using shared-exclusive locks, but we argue that Block Protection Protocol is better-suited for \sys{'s} graph workloads. %has better performance well-suited for \sys{'s} graph workloads. 
\sys{} already handles edge and vertex-level write-write conflicts using locking bits and $CAS$. Most of the time, an edge-deltas block is in $normal$ state, and a transaction $t$ only needs to read the state variable, and to store the block ID in its {\em block\_access\_table} entry. $t$ does not write to the shared memory variables to reduce cache invalidation across cores. Only during edge-deltas block consolidation and vertex deletion, $t$ needs to modify shared state variables and scans the {\em block\_access\_table} but these are rare.
%Though a worker thread needs to register its block access in the global $block\_access\_table$ (writing to global memory), it only needs to read other threads' entries during exclusive state changes in edge-deltas block consolidation and vertex deletion, which are much rarer than normal operations. Therefore $block\_access\_table$ (of other threads) is rarely accessed and unlikely stored in cache, avoiding cache invalidation. As a result, although ensuring mutual exclusion has become more expensive than using locks,  
%However, these operations are only needed for edge-deltas block consolidation and vertex deletion that are much rarer, and only the consolidation or vertex deletion thread needs to pay this extra cost.
We trade-off their extra costs for low overhead execution of the more common normal transaction operations.
%\vspace{-2mm}
\subsection{Consolidation of an Edge-Deltas Block}
\label{sec-consolidation}
\sys{} preallocates storage in edge-deltas blocks to reduce edge delta allocation overhead and provide sequential adjacency list access as in~\cite{openBwTree,LiveGraph}. As edge deltas build up, edge-deltas blocks become full. \sys{} needs to allocate new blocks to accommodate new edge deltas and recycle memory of the invalidated deltas. \sys{} edge-deltas block consolidation analyzes edges, and allocates a new block to store the latest version and future edge deltas. %latest versions, allocates a new block to accommodate existing and future edge deltas, and copies latest version edge deltas to the new block. 
\sys{} consolidation has 2 features: Reduce interference with concurrent transactions via the Block Protection Protocol, and adjust parallelism in the adjacency list to accommodate workloads using adaptive delta-chain locking. %The former is achieved through Block Protection Protocol, and the later is achieved through adaptive delta chain locking.
%We exhibit the block consolidation procedure using an example in Figure~\ref{fig:cons-example}. 
Refer to the example in Figure~\ref{fig:cons-example}.
Operations in each phase are in different colors.
V20's Edge-deltas Block $b$ of Size 512 bytes is full with a 64 byte header, 6 edge deltas (384 bytes) and 64 bytes of property data with {\em combined\_offset=(64<<32+384)}. Two ongoing transactions $T1$ and $T2$ 
%already created 
have
edge deltas in $b$.  Transaction $t$ updates an edge and %After it passes the write-write conflict check, 
%It 
increments {\em combined\_offset} to {\em (64<<32+448)} to allocate a delta. $64+ 448= 512$ is greater than the total space, so $t$ starts $b$'s consolidation. $t$ changes $b$'s state to {\em overflow}, waits till no other transactions are accessing $b$, and restores {\em combined\_offset} to {\em (64<<32+384)}. %The reason of this move is that the overflow $combined\_offset$ is corrupted and cannot locate where edge deltas start. 
The corrupted overflow {\em combined\_offset} cannot locate where edge deltas start. Resorting it in an exclusive state prevents concurrent transactions from accessing it. The atomic {\em fetch\_add} 
%used for memory allocation 
ensures 
that 
only 1 transaction causes the overflow. 

{\noindent \textbf{Analysis:}} $t$ changes the state to $consolidation$, and analyzes $b$. %Concurrent readers are allowed to access $b$ and transactions can update edge delta timestamps for commit and abort. 
Only edge updates are prohibited in this state. $t$ scans $b$ and finds latest version edge deltas: $d(4,UPD)$, $d(5,INS)$, and $d(3,INS)$. It also records $d(5,UPD)$ of $T1$ and $d(4,DEL)$ of $T2$ as in-progress. $t$ calculates a new block size using the workload history: latest version and in progress edge deltas. 
%It needs at least 418 bytes for the new block: 
The new block needs at least 418 bytes:
352 bytes to accommodate latest version and in progress deltas, and 64 bytes for the block header. %setting a lower bound of  $352+64=418$ bytes. 
Thus, $t$ allocates a new block of size 1024 bytes; the smallest power of 2 that is at least 1.5x of 418, and 4 delta chains. \textcolor{black}{We design multiple different heuristics to control the adaptivity of edge-deltas blocks. One heuristic {\em Simple} 
%simply 
calculates the smallest power of 2 of the minimum required delta storage (e.g., 512) here, and checks the new block's fill ratio. If the ratio is large, e.g., larger than $75\%$, it will promote the block size to the next power of 2 (e.g., 1024) to avoid repetitive consolidations. Another heuristic {\em LifeSpan} calculates the ratio based on the distribution of creation and invalidation timestamps of all deltas in the edge-deltas block, and how long the block exists between its creation time and the consolidation time. A long-living block indicates coldness, and requires a smaller ratio, e.g., $1\times$, and a short-living block indicates hotness and requires a larger ratio. 
%For our experiment we found the {\em Simple} 
In the experiments, it is found that {\em Simple}
already suffices. We aim to develop more sophisticated models and functions to calculate the ratio for the {\em LifeSpan} heuristic 
%as a future work.} 
in the future.}
%\sys~ uses a heuristic to calculate the ratio, e.g., 1.5x, based on the delta timestamps for adaptivity (Details to be given in~\cite{anonymous-gtx-tech-report} due to space).
%WGA: This is unclear. Whaty do you mean by 2 in-progress deltas and delta timestamps?
%With a larger block size, 
%Libin: I see. I was trying to say we use a heuristic, taking into the account that there are 2 concurrent transactions writing deltas, and the timestamps in the current block, to derive a ratio number 1.5. This shows the adaptivity. But I don't want to go too much into the details here. Like explaining the full heursitic.
%WGA:Adaptivity is an important feature. However, determining the size of expansion is a detail that can be omitted, I believe. It is good to say that the size can adaptively grow (or shrink, I assume), but by how much, I believe this can be deferred to the anonymous tech report. What do you think?
%Libin: I think we can say something like this. Let me draft one.
%Larger block sizes can make edge delta chains accommodate more edge deltas, and hence enable more concurrent updates. 
Larger block sizes and more edge delta-chains can accommodate more edge deltas, and enable more concurrent updates.
$t$ allocates a new edge-deltas block $b'$ and updates its header: pointing $b'$'s previous pointer to $b$, and setting $b'$'s creation timestamp to 10; 
%\walid{Does the two "its" below refer to t or to b? Please fix.}%Libin: fixed
%Professor: can you please explain why you changed ":" to ";" for "creation timestamp to 10"? I want to indicate 10 is the largest invalidation timestamp. %WGA: I believe using ; fits better what you intend to convey, i.e., to explain why. If you do not like it, change it to : but it would be unusual in my opinion.
the largest invalidation timestamp in the $b$. Transactions with $rts<10$ can access $b$. %The creation time also serves as the watermark to recycle $b$. 
$t$ also registers that $b$ can be safely recycled after Timestamp 10.
$t$ writes the 3 latest version edge deltas to $b'$, and updates the delta-chains index. 

{\noindent \textbf{Installation:}} After installing all latest version edge deltas, $t$ changes the block state to $installation$ to synchronize %between in progress 
transactions. \sys{} handles concurrency at delta chain level that may change after a consolidation. %allocate a new edge-deltas block with a different number of delta chains. 
To avoid inconsistencies in delta chains after consolidation, $installation$ synchronizes committing transactions before swapping the edge-deltas blocks. %In section~\ref{sec-edge-update}, 
$t$ updates delta-chains index entries to ``install" its edge deltas during commit (Section~\ref{sec-edge-update}). $installation$ requires all transactions that have already ``installed" their edge deltas in $b$ to commit or abort before $b$ continues %starts %Libin: I feel continue is a better word to describe the situation.
consolidation, while committing transactions cannot install their edge deltas if $b$ already has reached $installation$. $t$ knows which delta chains and transactions to check based on the previous analysis. %The commit protocol is covered in the next section~\ref{sec-hybrid-commit}. 
Assume that $T2$ has  installed its edge delta but later aborts and $T1$ is still running. %executing its operations. %, and $T2$ aborts. 
$t$ waits till seeing $T2$'s final status(abort), and only writes %$T1$'s edge delta 
$d(5,UPD)$ to $b'$. %the new block. %$b'$'s delta-chains index is not updated for this in progress edge delta. 
If $T2$ were to commit, $t$ would also write $d(4,DEL)$ to $b'$, and update the delta-chains index to point to $d(4,DEL)$. %reflect it. 
%\walid{"it" refers to what? the update?} %Yes, to reflect the commited edge delta. I have modified the statement, do you think this is OK?
After 
%the 
installing all in-progress edge deltas, $t$ increases the block version number, swaps $block\_ptr$ and index, % to $b'$ and the new delta-chains index
and 
%puts 
switches
the block state back to $normal$. It allows other transactions, 
%like 
e.g., $T1$, to access the block. Later, $T1$ commits and updates its edge delta at $cts=18$. Using the Block Protection Protocol ensures little overhead when the block is not overflowing, and improves concurrency during consolidation. Block consolidation adjusts block size and concurrency level to adapt to the workload: Hot(cold) blocks allocate more(less) memory and provide more(less) concurrency.
%Block consolidation increases edge-deltas block size and delta chain number to accommodate more edges and concurrent updates, adapting its concurrency level to the workload. Note that the ``cold" block's size may also decrease, trading less concurrency for saving memory.
%\vspace{-3mm}
\subsection{Lazy Update}
\label{sec-lazy-update}

%%\vspace{-6mm}
\sys{} uses Lazy Update to reduce latency of transaction commits, and enforce SI~\cite{SI-ref,mvcc-ref}. 
It %maintains a distributed 
distributes a 
transaction table to store transaction status: Each thread maintains its own transaction table, and can 
%visit 
access
other transaction tables based on thread IDs. We use 
the most significant 8 bits in transaction IDs 
to
store worker thread information. Each transaction ID also has the most significant bit set to distinguish 
%with 
them from
timestamps as discussed below.
%\walid{Did you say at this point that you can either store TID or timestamp in the same location and that bit states which is which? and did you state why you do that? Do not recall seeing it. But if you did, please ignore this comment, or point the reader to the proper section where this is discussed.}
%\Libin{hello professor, yes I did. In vertex and edge update sections 5.1 and 5.2, I mentioned transactions store transaction IDs as timestamps of deltas and later update them to real timestamp after commits. In the introduction of section 6, I mentioned transactions store transaction IDs as timestamps of deltas to 1). be able to see its own writes, and 2). can determine an object is being updated by a transaction. But I never explicitly make a consolidated statement like "transactions update objects by creating deltas with their transactions IDs as timestamps, and later commit by updating them to commit timestamps." Should I try to make a statement like that?}
When a worker thread starts a new transaction, it calculates the transaction ID from its thread ID and a local counter, and stores the transaction entry in its transaction table. Each worker thread maintains every transaction $t$'s $status$ and $op\_count$ in its transaction table. %2 attributes for every transaction $t$ in its transaction table: $status$ and $op\_count$. 
$status$ records $t$'s current state: $In\_Progress$, $Abort$ that are predefined 64-bit integers or a positive commit timestamp. $op\_count$ records the number of deltas created by $t$. $t$ has $status=In\_Progress$ and $op\_count=0$ at creation time, and increments the $op\_count$ as it creates more deltas. If $t$ aborts, its $status$ is set to $Abort$. If $t$ commits, its $status$ stores its commit timestamp.
Figure~\ref{fig:txn-table} gives an example of Thread 5's transaction table. 
Thread 5 has created 17 transactions and the table contains its {\em 15-17}$^{th}$ transactions' information.

At runtime, a transaction $t$ may access an edge-deltas block or vertex delta but encounters another transaction $q$'s in-progress (edge) delta with $cts=q$. %another transaction, say $q$, with private {\em In\_Progress} edge-deltas. 
%As the next section~\ref{sec-hybrid-commit} shows, \sys{}'s group-commit manager commits transactions by updating their status in the transaction table to a commit timestamp.% without updating the transactions' deltas.
$t$ examines $q$'s status via $q$'s ID, and updates the delta's $cts$ %create timestamp 
(and previous version's $its$%invalidation timestamp
) to $q$'s $wts$ %commit timestamp 
if $q$ has committed. Then, $t$ decrements $q$'s $op\_count$ entry. Each transaction table is a circular array, and its entries can be reused if $op\_count$ equals 0. It  enables $O(1)$ lookup of transaction status using transaction ID.
Similar approaches for managing delta commit, abort, and visibility by cooperative threads are used in CTR's fast recovery~\cite{constant-time-recovery-ref}, Hekaton's cooperative resource management~\cite{hekaton-ref}, and Hint Bits for reduced visibility check overheads~\cite{hint-bits-ref,hints-bits-blog-ref}. 
%Libin Sep 5th: I shrink this part a bit.
%CTR stores transaction abort status in a global Aborted Transaction Map (ATM), and allow transactions to start immediately after recovery's redo phase. Concurrent transactions examine transaction status and undo updates from aborted transactions during runtime. Hekaton's index scanners unlink garbage during index operations. PostgreSQL's hint bits allow concurrent transaction to mark versions as visible to all transactions to speed up visibility checks: preventing transactions from excessively consulting ($pg\_clog$). The first transaction that finds the transaction commit status in the $pg\_clog$ updates the version's hint bits. 
%These protocols provide fast recovery~\cite{constant-time-recovery-ref}, cooperative resource management~\cite{hekaton-ref}, and reduced visibility check overheads~\cite{hint-bits-ref,hints-bits-blog-ref}. %transaction execution with less overhead~\cite{hint-bits-ref,hints-bits-blog-ref}.
%However, as we will show in the next subsection, 
\sys{} uses Lazy Update for a new purpose, namely, high throughput and low latency transaction commit.
%\vspace{-3.5mm}
\subsection{Hybrid Transaction Commit and Abort}
\label{sec-hybrid-commit}
%\Libin{todo: add discussion of version validation}
\sys{} designs a high throughput low latency commit protocol for its read-write transactions inspired by Constant Time Recovery (CTR)'s fast recovery using a global table~\cite{constant-time-recovery-ref}. %The high performance recovery protocol CTR~\cite{constant-time-recovery-ref} implements a \textbf{constant time} recovery to make database available right after the redo phase. 
%For this goal, 
%We learn from Constant Time Recovery(CTR)'s fast recovery using a global map/table~\cite{constant-time-recovery-ref}.
\sys{'s} hybrid group-commit protocol combines Lazy Updates and group commits to support fast transaction commit while maintaining SI correctness~\cite{SI-ref,mvcc-ref}. 
\sys{} has a commit manager thread that manages a global commit array, where each worker thread has an entry to register its committing transaction.
A committing transaction $t$ starts with a preparation phase: $t$ makes its edge deltas the heads of 
their 
corresponding edge delta-chains, and unlocks them by updating the associated delta-chains index entries (Sections~\ref{sec-edge-update},~\ref{sec:delta-chain-locking}).
Section~\ref{sec-consolidation} introduces consolidation that may modify delta chains storage. Thus, during the preparation phase, $t$ must register block access using the Block Protection Protocol, and checks block version. $t$ is only allowed to install its edge deltas in $normal$ and $consolidation$ states. For other states, $t$ has to wait and retry. If the block version has changed since $t$ wrote its last delta, $t$ has to ``relock" all delta chains its edge deltas belong to. Only after locking all needed delta chains, $t$ can install its edge deltas. Otherwise, $t$ aborts. $t$ executes commit preparation in a vertex ID order to avoid deadlocks.
After $t$'s preparation phase, future transactions can lock these delta chains, and lookup $t$'s edge deltas using the delta-chains index. 
%Such 
These
operations are not needed for vertex deltas.
Then, $t$ puts a pointer to its transaction table entry in its commit array entry.
%When the array is not empty, the commit manager adds 1 to $gwe$, uses the return value as the group commit timestamp ($wts$), and scans the commit array. 
The commit manager adds 1 to $gwe$ as the group commit timestamp ($wts$), and scans the commit array.
For each 
%transaction (e.g., $t$) 
transaction, say $t$,
in the array, the commit manager writes $wts$ as $t$'s status in the transaction table entry. %, and removes it from the array. 
%$t$ has committed at $wts$. 
After scanning the whole array, the commit manager {\em fetch\_add}s  $gre$ by 1, \textbf{atomically committing} all transactions in this commit group at $wts$. After being assigned a $wts$, $t$ eagerly updates its deltas' timestamps while the concurrently executing transactions can Lazy Update them, making up the hybrid commit protocol. 
Correctness-wise, transactions starting after %the commit manager increases 
the incremented $gre$ are guaranteed to see deltas of the commit group. Performance-wise, it allows a group of  transactions to commit without updating all their deltas, and the new $gre$ to start immediately. It reduces group commit latency and allows the commit manager to work on the next commit group without waiting for the previous group to commit all its deltas. It reduces  potential stalls of the new commit group caused by transactions with large write sets, %: if the current commit group contains transactions with a large write set, they may take a long time to update all their deltas, and become the bottleneck in the group commit.
and allows concurrent transactions to cooperate and parallelize these transactions' delta commits. %, automatically parallelizing delta commits. 
It also decreases the synchronization between the commit manager and the commit group. 
%\textcolor{red}{Although the similar cooperative management of versions exist for state-of-the-art systems, they aim for different tasks including recovery~\cite{constant-time-recovery-ref}, visibility management~\cite{hint-bits-ref,hints-bits-blog-ref}, and garbage collection~\cite{hekaton-ref}. \sys{'s} main focus is on high throughput transactions, and we design the commit protocol for this purpose. Later in the experiment section, we will exhibit \sys{'s} superior transaction throughput. }\Libin{I feel my argument is slightly weak. I know I have arguments in mind for these related works but not doing a good job presenting it yet.}

{\noindent{\bf Transaction Abort}}: 
%When a read-write transaction $t$ aborts,  
%due to writ-write conflicts. Compared to committing, a transaction can abort by 
%it manually aborts all its deltas. 
An aborting read-write transaction $t$ writes $Abort$ to its transaction table entry. 
Then, $t$ iterates over its modified edge-deltas blocks, 
%registers access, 
%and 
aborts all of $t$'s deltas by writing $Abort$ to their creation timestamps, and unsets any lock bits in the delta chains. $t$ aborts all its vertex deltas by removing them from the delta chains.

%\vspace{-3mm}
\subsection{State-aware Transaction Operations}
\label{sec-state-aware-operations}
\noindent{\bf Edge Operations.} Transactions use %register access and pass 
the Block Protection Protocol for edge operations: states $normal$ and $consolidation$ for reads, and $normal$ for writes. To accommodate block version change after consolidation, readers compare their $rts$ with blocks' creation timestamp, and may need to use previous pointer to read an earlier edge-deltas block. %Writer transactions need to detect write-write conflicts using lock bits. 
%They can execute edge lookup and adjacency list scan only if the block state is $normal$ or $consolidation$. Writer transactions can register access only at $normal$ state and they still need to detect write-write conflicts using lock bits. 
%When accessing a block that writer $t$ has edge deltas in, $t$ has to check whether the version number has changed.
%When a writer $t$ accesses a block that $t$ has edge deltas in, and the block version number has changed, 
%If so, 
%$t$ has to relock delta chains as in the commit preparation phase in section~\ref{sec-hybrid-commit}. $t$ will abort if it fails to relock any delta chains. 
Writers relock delta chains as in the commit preparation phase (Section~\ref{sec-hybrid-commit}) if they have edge deltas in the block with changed version and will abort if the relocking fails.

\noindent{\bf Vertex Delete.} Block Protection Protocol has a an exclusive state $delete$ for deep cascading vertex deletes at high priority. %Assume transaction $t$ is deleting vertex V1. %It installs a special delete delta as V1's new version. %following vertex updates in section~\ref{sec-vertex-operations}. 
For example, when deleting V20 (Section~\ref{sec-vertex-deletion}), $td$ sets each of V20's edge-deltas block's state to $delete$ to delete the whole adjacency list. %waiting for threads that are accessing the block to exit.
%Following the discussion on deleting V20 in section~\ref{sec-vertex-deletion}, $td$ sets V20's each edge-deltas block's state to $deletion$ waiting for threads that are accessing the block to exit. %$deletion$ is an exclusive state so $t$ has to wait for threads that are accessing the block to exit. 
Concurrent transactions will \textbf{abort} if they see $delete$ on an edge-deltas block. % they want to access. %This design gives vertex deletion a high priority for a timely deletion. %Then $td$ follows the procedure in section~\ref{sec-vertex-deletion}.%The other parts of vertex deletion follows the procedure in section~\ref{sec-vertex-deletion}. 
After deleting all V20's edges in neighbors' edge-deltas blocks, $td$ follows the same hybrid commit protocol. %commits similarly as a normal transaction in the hybrid commit protocol. %However, when deleting edges, $t$ can hold resources and wait for delta chain locks instead of aborting.%we allow $t$ to hold resources and wait for locks instead of immediately aborting. 
One difference is that $td$ never aborts, and  waits to lock delta chains to delete reverse edges of V20.
\textcolor{black}{After $td$ has committed, it registers V20's vertex delta and all its edge-deltas blocks safe to be recycled after $td$'s commit timestamp (more details in Section~\ref{sec-resource-management}).}
%We give 
\sys{} assigns vertex delete  highest priority to satisfy privacy laws like European Union’s General Data Protection Regulation that requires timely deletion of user data~\cite{ldbc-snb-bi-ref,ldbc-snb-int-v2-ref,gdpr-impact-ref}.
\section{Durability and Recovery}
\subsection{\textcolor{black}{Group Redo Logging}}
\textcolor{black}{Since both \sys{} and LiveGraph~\cite{LiveGraph} 
%both 
use group commit, we extend \sys{'s} commit protocol by  redo logging
that is similar to that of LiveGraph. Each transaction maintains a redo logging buffer in its transaction table entry, recording newly created vertex and edge versions. After the commit manager scans the commit array and determines the commit group, it groups all individual transactions' commit logs into blocks and writes them to disk. \sys{} only maintains redo logging because there is no persistent store to be undone, similar to other main memory DBMSs~\cite{main-mem-recovery-ref,hekaton-ref,silor-ref}. It also adopts value logging as in Silor~\cite{silor-ref} to enable highly parallel recovery. We support both synchronous and asynchronous logging. 
%At default 
\sys{} 
%\walid{What do you mean by: At default?}
%\Libin{Later in the section I said we also have an option to do asynchronous group logging which has better performance but may lose transaction updates. Here I was trying the default setting is synchronous (forced flush).}
forces 
the 
group commit log to be persisted before committing the transaction group. Transactions can 
%eager and lazy 
eagerly or Lazily
update deltas but those deltas 
%won't 
would not 
be visible to new transactions until 
the
commit manager persists the log and updates the $gre$. It ensures full durability but limits performance. \sys{} also supports asynchronous logging~\cite{main-mem-recovery-ref}. It allows transactions to commit, and a new commit group to start while their logs are still buffered in memory (later written to disk in larger blocks). It has less influence on the normal transaction throughput at the risk of losing committed transaction updates. Our experiments show $20\%$ and $12\%$ throughput degradation for synchronous and asynchronous logging, respectively.}
\subsection{\textcolor{black}{Incremental Fuzzy Checkpointing}}
\textcolor{black}{Inspired by the main-memory systems Hekaton~\cite{hekaton-ref} and Silor~\cite{silor-ref}, we design a delta-based fuzzy checkpoint protocol that is collaborative, parallel, and incremental. \sys{} manages a global checkpoint object that is accessible to all worker threads. It contains the timestamps the current and last checkpoints begin ($current\_ts$ and $last\_ts$), the maximum vertex ID $max\_vid$ when this checkpoint begins, and the current checkpoint progress $current\_vid$. %, and some additional atomic flags to start and end the current checkpoint. 
After $x$ transaction operations, Worker Thread $t$ enters the checkpoint phase. If the checkpoint is ``uninitialized", $t$ initializes it by updating atomic flags. It stores \sys{'s} maximum vertex ID, and $gre$ in the checkpoint's $max\_vid$ and $current\_ts$, and creates a new checkpoint directory at a system-designated location for all checkpoints. Each checkpoint directory stores the $current\_ts$ in its name to maintain a chronological order of checkpoints. % with a file storing the last checkpoint directory address. 
Each worker thread appends to a separate checkpoint file within this directory. After initialization, or if the checkpoint is already initialized, $t$ 
%does 
performs
one round of cooperative checkpointing. $t$ $fetch\_adds$ the $current\_vid$ by a number $z$. Assuming the return value is $k$, $t$ checkpoints Vertices $k$ to $k+z-1$. %For each vertex, $t$ reads its 
%\walid{"its" refers to the vertex. Correct?}
%\Libin{Yes it does, t reads v's version.}
For each vertex, say $v$, $t$ reads $v$'s
latest version vertex and edge deltas from the vertex delta chain and edge-deltas blocks. All \textbf{latest version deltas} with $cts>last\_ts$ will be written to $t$'s checkpoint file sequentially. %For each edge-deltas block, $t$ also records how large the current block is to enable appropriate allocation during recovery. 
Also, $t$ 
%also 
records each edge-deltas block size to enable appropriate allocation during recovery.
After $current\_vid>max\_vid$, the current checkpoint finishes. The last worker thread to finish its local checkpoint updates the system metadata to 
%make 
mark 
the new checkpoint directory 
as 
the latest checkpoint. 
\sys{'s} checkpoint is fuzzy. It 
%doesn't 
does not
guarantee a transaction-consistent state: a transaction started after $current\_ts$ may only have parts of its updates persisted in the snapshot. We choose this design for the same reasons as those for  Silor~\cite{silor-ref}. We 
%don't 
do not
want the system to force a snapshot on checkpoint which may delay garbage collection and incur extra memory allocation cost. Also \sys{} has to replay redo logs starting from $current\_ts$, so there is no need to enforce a transaction-consistent snapshot. \sys{} checkpoint is incremental. A new checkpoint stores only the \textbf{latest version} changes since the last checkpoint. \sys{} checkpoint files are append- and then read-only. When the checkpoints reach a threshold, an LSM-tree-styled merging will merge the disk files into a single checkpoint.
%(though this checkpoint is still stored in multiple files). 
Our experiments show an additional {\em 5-10\%} throughput degradation depending on the checkpoint frequency than only using the redo log while reduces the log size by up to {\em 95\%}.}
\subsection{\textcolor{black}{Parallel Recovery}}
\textcolor{black}{%We want \sys{} to have the same high parallelism in recovery. like Silor and Hekaton~\cite{silor-ref,hekaton-ref}. Therefore we choose value logging, and store checkpoints in multiple files. 
We propose a parallel recovery 
technique 
%like 
similar to those of 
Silor and Hekaton~\cite{silor-ref,hekaton-ref} using value logging and multiple checkpoint files to have high parallelism in recovery.
Within each checkpoint directory, each checkpoint file stores a disjoint set of vertices and their edges. 
\sys{} recovery starts with the latest checkpoint directory. Each worker thread loads its own checkpoint file in the directory, and writes 
the
latest version edges and vertices into \sys{}. This step has little interference among threads because each checkpoint file stores a disjoint set of vertices and edges. After finishing one checkpoint, worker threads move on to the previous checkpoint until all checkpoints are processed. Then, worker threads replay the logs from the log tail until the latest checkpoint's timestamp. Worker threads replay operations in parallel 
%like 
as in
normal edge and vertex 
%update 
updates
but create deltas with actual creation timestamps instead of transaction IDs. Notice that for every vertex or edge in a checkpoint and redo logging,  reconstruction starts with locating the previous version delta. The replay takes place only if the operation's create timestamp is larger than the latest version delta's.}
\section{Serializable Transactions}
\textcolor{black}
{\textcolor{black}{Serializable read-write transactions} are implemented using read set validation~\cite{silo-ref,hekaton-ref,cicada-ref}. While  \sys{'s} SI only requires transactions to read a consistent snapshot, Serializability in \sys{} requires a transaction's reads and writes to logically happen at the same time (its read-set stays unmodified during transaction execution)~\cite{hekaton-ref}. In $MVCC$ systems, read-only transactions acquire a read timestamp ($rts$) at create time, and sees all versions with overlapping lifespan. These read-only transactions can always be put in a serial order according to their $rts$. Thus, our focus is on read-write transactions with both reads and writes on the graph. }

\textcolor{black}{Each read-write transaction, say $t$, locally keeps a read set $R$, scan set $S$, and a write set $W$. $t$ stores all vertices and edges it reads in $R$, and all adjacency lists it scans in $S$. $t$ records each of its writes in $W$. When $t$ searches an edge delta-chain or scans an edge-deltas block, $t$ checks whether the edge delta-chain or edge-deltas block contains a delta with larger $cts$ than $t$'s $rts$. If this is the case, $t$ can abort early because $t$'s writes will break serializability. When $t$ is committing, besides the preparation phase to install edge deltas (Section~\ref{sec-hybrid-commit}), $t$ has to validate each vertex/edge in its read set, and edge-deltas block in its scan set. $t$ groups edges in its read set and edge-deltas block in its scan set based on vertex IDs and labels, and validates them in this order. It further groups edges in the same edge delta-chain together. For $v,e\in R$, $t$ checks the vertex or edge delta-chain head's $cts$. If $cts>t.rts$, $t$ aborts because its read set has already been updated. For $t$'s scan set, $t$ has to validate that the edge-deltas block (adjacency list) does not have new edge deltas created after $t$'s $rts$. Otherwise, $t$ aborts. After $t$ validates its read and scan sets, it can move forward to commit. We revise the commit phase with additional synchronizations. Each transaction must register in the commit group to start its validation phase. After the validation phase, 
%it 
the transaction
fetches the global write epoch, and uses it as its commit timestamp. To accommodate the validation phase, the commit manager tracks the current commit group, and enforces a serialization point between groups. It periodically enters a quiescent state and waits for all transactions in the current commit group to validate, then commit or abort. It starts a new commit group by incrementing global read/write epochs. 
Future transactions must wait to commit after the quiescent state. Note that if $t$ has an empty write set or it contains empty read and scan sets, it is considered a read-only or a write-only transaction, and does not participate in read set validation.}

{\noindent}\textcolor{black}{{\textbf{Discussion:}} There are some implications 
that follow
from designing serializable transactions. If a transaction $t2$ is placed after transaction $t1$ in the serial order according to the commit timestamp, $t2$ should see all $t1$'s committed updates. Therefore, $t2$ has to ensure that it does not miss any updates. If $t2$ repeatedly fails the validation when it reads a hotspot edge-deltas block, a lock-based protocol is preferred to ensure progress. On the other hand, using a read-write transaction to perform both graph analytics and updates will significantly degrades performance.
%have catastrophic consequences. 
%\walid{"catastrophic" is a subjective word. What do you mean by it? Be objective or quantitative.}
%\Libin{I meant to say the performance will suffer significantly under this situation. As I argued later, a single concurrent write will abort this long-running transaction under serializability. Shall we change it to something like "high risk or abort" or "significantly degrades performance"?}
Because graph analytics algorithms typically scan the whole graph, a single update will break the serializability for this transaction if it also has writes. Therefore, it is better to use a read-only transaction for graph analytics. If there is a need to run a transaction with both graph analytics and updates, it is better to reduce the isolation level to Snapshot Isolation instead.}
%\vspace{-2mm}
\begin{table*}[htbp]
    \centering
    \begin{tabular}{|c|c|c|c|c|c|c|c|}
        \hline
         \textbf{Systems}&\multicolumn{2}{c|}{\textbf{Atomicity}} & \multicolumn{2}{c|}{\textbf{Consistency}} & \multicolumn{2}{c|}{\textbf{Isolation}}&\textbf{Durability}\\
         \hline
         {}&single operation & multiple operations & no dangling edge & no duplicate edge &SI & Serializable & log+checkpoint  \\
         \hline
         \sys{} & \cmark & \cmark & \cmark& \cmark & \cmark & \cmark & \cmark \\ 
         \hline
         LiveGraph & \cmark & \cmark & \cmark& \cmark & \cmark & \xmark & \cmark \\
         \hline
         Teseo & \cmark & \cmark & \cmark& \cmark & \cmark & \xmark & \xmark \\
         \hline
         Sortledton & \cmark & \cmark & \cmark& \cmark & \xmark & \cmark & \xmark \\
         \hline
         Spruce & \cmark & \xmark & \xmark& \cmark & \xmark & \xmark & \xmark \\
         \hline
    \end{tabular}
    \vspace{5mm}
    \caption{Transactional Graph System Comparison}
    %\vspace{-4mm}
    \label{tab:system-comparison}
\end{table*}
\begin{table}[htbp]
    \centering
    \begin{tabular}{|c|c|c|c|c|c|c|c|}
        \hline
         \textbf{Graph Dataset}&{\textbf{|V|}} &{\textbf{|E|}} &{\textbf{Avg. Degree}}&\textbf{Max Degree}\\
         \hline
         {\em dota-league} & 61K & 51M & 1663.24 & 17004\\ 
         \hline
         {\em graph500-24} & 9M & 260M & 58.7 & 406416\\ 
         \hline
         {\em uniform-24} & 9M & 260M & 58.7 & 103\\ 
         \hline
         {\em yahoo-songs} & 1.6M & 256M & 513.04 & 468425\\ 
         \hline
         {\em edit-wiki} & 50M & 572M & 22.561 & 5576228\\ 
         \hline
         {\em graph500-26} & 33M & 1B & 64.13 & 1003338\\ 
         \hline
         {\em twitter} & 41M & 1.46B & 70.506 & 3081112\\ 
         \hline
    \end{tabular}
    \vspace{2mm}
    \caption{{Dataset Statistics}}
    %\vspace{-4mm}
    \label{tab:dataset}
\end{table}
\section{Resource Management}
\label{sec-resource-management}
\noindent{\bf Memory Allocation.}
%Libin: Sept 23rd: if the memory allocator is directly adopted from livegraph, maybe simplify it even more?
\sys{} adopts the hybrid memory allocator from LiveGraph~\cite{LiveGraph} to balance concurrent memory allocation and space utilization. It allocates a large chunk of memory at system start via $mmap$, and divides memory into blocks of varying powers-of-2 sizes (as in Buddy systems~\cite{buddy-ref}). Each worker thread locally manages a pool of memory blocks smaller than a threshold and can allocate memory blocks from this pool without synchronization. %without synchronizing with the other threads. 
Larger memory blocks are managed %by a memory manager 
in a global pool. % that handles requests to allocate large memory blocks. 
Due to the power-law nature of real-world graphs, most vertices are not hub vertices, i.e., do not have many edges. Thus, most of the vertex delta and edge-deltas block can be directly allocated from threads' private pools. %'s memory allocations only require smaller blocks from threads' private pools. 
Even when threads need to allocate a larger block, they unlikely request it at the same time. 
Thus, \sys{}'s memory allocation protocol is mostly contention-free.
%yet at times requires the memory manager to allocate large memory blocks. 

\noindent{\bf Cooperative Garbage Collection.} $MVCC$ creates garbage: versions that no current and future transactions can see. State-of-the-art $MVCC$/snapshot systems support garbage collections (GC)~\cite{silo-ref,high-performance-mvcc-ref,hekaton-ref,cicada-ref,constant-time-recovery-ref,LiveGraph,Teseo,Sortledton,spruce-ref}. Cooperative garbage collection is used extensively in both relational and graph systems due to its good performance in coupling with normal transaction operations~\cite{LiveGraph,Sortledton,silo-ref,high-performance-mvcc-ref,hekaton-ref,cicada-ref}. It saves the overhead of maintaining additional garbage collection threads, and provides  timely recycling of garbage. %A transaction can remove a garbage at the first encounter without waiting for the service thread to pick it up.
%\sys{} recycles the delta blocks no longer visible to transactions. 
%\sys{} is cooperative with no garbage collector thread. Instead, 
\sys{} maintains a timestamp table (one entry per worker thread) to store the timestamps of running transactions.
Worker threads register memory blocks as garbage during runtime, and periodically recycle the ``safe" blocks.  When a transaction commits or Lazy Updates a vertex delta, it 
\textbf{invalidates} the vertex's previous version.
A worker thread invalidates the (old) edge-deltas block after consolidating it. %at the new block's create time. 
 %indicates that the previous version of the vertex is invalidated.
 %at the same time. 
Each worker  thread, say $w$, maintains a local priority queue to store the invalidated vertex delta or edge-deltas blocks with their invalidation times. 
%\textcolor{red}{Silo~\cite{silo-ref} and Cicada~\cite{cicada-ref} have a similar garbage registration for its records when a record is deleted or a new version is created. The worker thread registers the previous version in its thread-local list/priority queue according to invalidation timestamps.}
%After executing a certain amount of transactions, 
Periodically, $w$ scans the timestamp table to find the ``safe" (minimum) timestamp. 
%currently stored 
%in the table. 
%It indicates all current and future transactions will have equal or larger read timestamps than this value; thus it is safe to clean all blocks invalidated before this timestamp. 
$w$ continuously %checks its local garbage queue, dequeues an entry, and 
recycles memory blocks from its garbage queue until the queue is empty or the current queue head's timestamp becomes too large.
%Note that memory block recycling is the reverse of the hybrid allocation. 
If a recycled block $b$ is small, $w$ %"rewards" itself by storing 
stores $b$ in its local pool, else $b$ is sent back to the global pool~\cite{LiveGraph}. %memory manager. 
This ensures that small blocks are allocated contention-free while large blocks return to the global pool for reuse by all threads. %~\cite{LiveGraph}.
%Libin: rewrote and shrink this part. The original text is in the commented paragraph following
Silo~\cite{silo-ref}, Hekaton~\cite{hekaton-ref}, and Cicada~\cite{cicada-ref} also register garbage, and perform cooperative safe timestamps-based garbage collection. \sys's GC is unique in its low overhead design. \sys{} recycles each outdated edge-deltas block as a whole instead of each invalidated edge delta individually, reducing garbage registration and recycling overhead. Each garbage registration is included in the normal operations: commit, Lazy Update or block consolidation. Thus, worker threads 
%don't 
do not
search for garbage. %\textcolor{red}{like Live~\cite{LiveGraph}}. 
\sys{} worker threads have little interference when recycling garbage blocks. Each garbage priority queue is managed locally, and the timestamp table is cache-friendly and latch-free. \sys{} GC ensures that the safe memory block is inaccessible by other threads.
Thus, there is no need for thread synchronization during GC.
In contrast, state-of-the-art GCs generally execute at per-object granularity~\cite{silo-ref,cicada-ref}, or interact with other threads through locks or atomic primitives~\cite{hekaton-ref,cicada-ref}.

\begin{figure}[htbp]
    \centering
    %\vspace{-4mm}
    \includegraphics[width=0.3\textwidth]{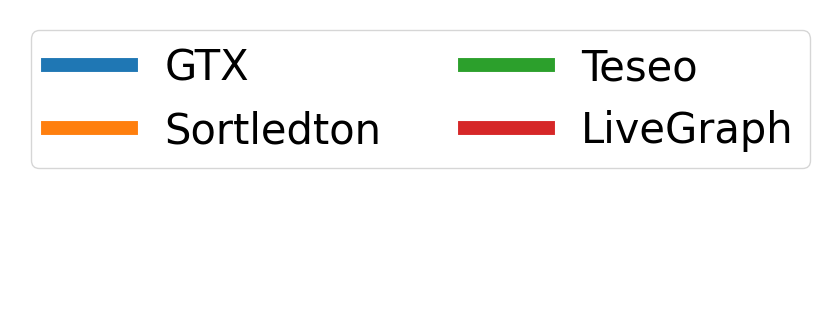}
    %\vspace{-10mm}
    \caption{\em Bars Color Code in Experimental Results. }
    %\vspace{-4mm}
    \label{fig:legends}
\end{figure}

\section{Evaluation}%Libin: sept 20th: need 1 more round of reviewing this section.
\label{sec-experiment}
%We present experiment results of Bw-Graph against other transactional graph libraries/systems. 
This study focuses on supporting high throughput read-write transactions with concurrent graph analytics.
\begin{figure*}[htbp]
    %width=1\textwidth
    \includegraphics[width=1\textwidth]%{Figures/Matplotlib_Figures/insertion/insertion_all.png}
    {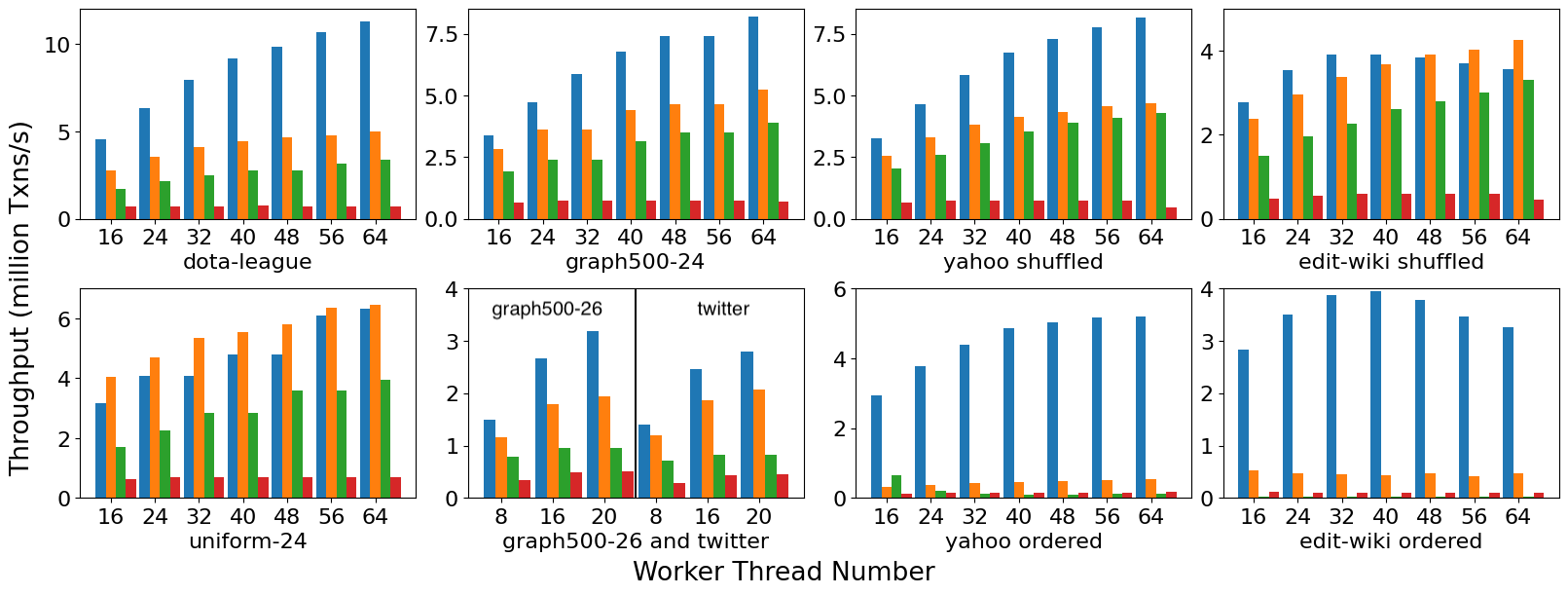}
    \caption{\textcolor{black}{Insertion Throughput} 
    %\wga{is color correct? check the bars figure code}
    } %\Libin{It is not. Sorry I made this mistake. Fixing now.}
    \label{fig:insertion}
    %\vspace{-2mm}
\end{figure*}
\begin{figure}[htbp]
    \centering
    \includegraphics[width=1\linewidth]{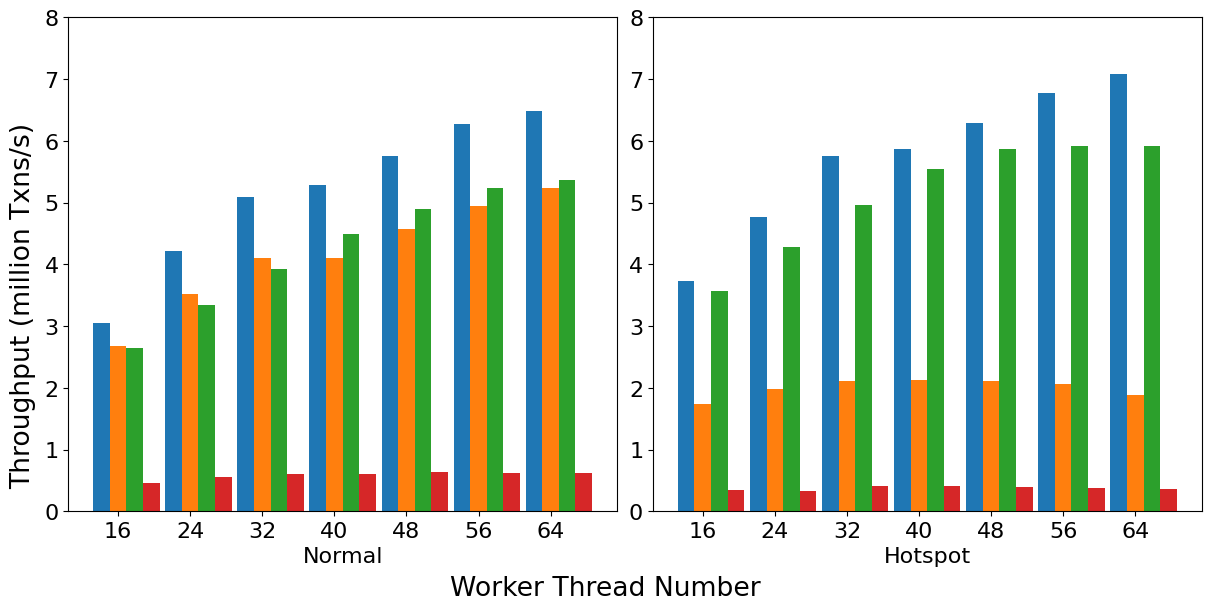}
    \caption{\textcolor{black}{Update Throughput}}
    %\vspace{-4mm}
    \label{fig:update-throughput}
\end{figure}
\noindent{\bf Experiment Setup.}
%Libin: I can also remove the experiment that we could not run on pm machine but on bigdata
Experiments run on a dual-socket machine with Intel(R) Xeon(R) Platinum 8368 @ 2.40GHz processors. {\color{black}It has 2 NUMA nodes and each NUMA node's CPU has 38 physical cores supporting up to 76 threads.} %with 156 CPUs. 
%\wga{Is it 152 not 156?}
%\Libin{Yes just confirmed}
%Each has 2 38-core NUMA nodes and around 96GB  RAM. 
%It has 2 NUMA nodes and each node has 38 cores and 96GB RAM.
%\wga{Is it 39 not 38?}
%\Libin{just checked, it is 38}
%It has 3.6 MiB L1 data cache, 2.4 MiB L1 instruction cache, 95 MiB L2 cache and 114 MiB LLC. 
We compile all systems with GCC {\color{black}13.3.0} and O3 optimization flag, and evaluate them over a single NUMA node with best effort, i.e.,
%to exclude NUMA-based performance issues.
%\Libin{Here I mean if the memory fits in a single NUMA node. The later section can optionally add more graphs}
%this one
the evaluated system allocates memory locally, and uses remote memory only if its memory exceeds a single NUMA node. 
{\color{black}We conduct experiments with up to 64 worker threads. %for the insertion experiment and up to 60 for the concurrent updates and analytics experiment. 
We choose to allocate fewer worker threads than one NUMA node's available threads to avoid causing computing resource contentions and induced inaccuracies and allow systems, e.g., Teseo~\cite{Teseo} to spawn service threads.%. We choose up to 64 worker threads to allow system like Teseo~\cite{Teseo} to spawn service threads.
} One exception is for insert experiments of {\em graph500-26}~\cite{graphalytics-data-ref} and {\em twitter}~\cite{konect-ref} with over 1 billion edges. This machine's memory is not enough to evaluate all the systems. 
%Libin: professor do you think if the below will a weak point?
%\wga{Do you evaluate all the experiments there in this case, or do you migrate only the ones needing more memory?}
Thus, for these datasets, we evaluate all systems in a setup with Intel(R) Xeon(R) Platinum 8168 CPU @ 2.70GHz processors that has {\color{black}12 cores supporting up to 24 threads} and 387GB DRAM per NUMA node. To minimize NUMA effects, we run the experiments in a single NUMA node with up to 20 worker threads to avoid {\color{black}computing resource contention}. %We disable logging. Libin: 1 CPU is div
{\color{black}Statistics of all graph datasets used in the experiments are given in Table~\ref{tab:dataset}.}
We use the same evaluation program based on the $LDBC$ Graphalytics benchmark~\cite{graphalytics-paper,graphalytics-specification} used in Teseo~\cite{Teseo}, Sortledton~\cite{Sortledton}, and Spruce~\cite{spruce-ref}, and add more experiments. We evaluate against the transactional graph systems: LiveGraph~\cite{LiveGraph}, Teseo~\cite{Teseo}, and Sortledton~\cite{Sortledton}. {\color{black}Our evaluation  focuses on SI isolation level, and we mainly evaluate the SI version of \sys{}. Note that LiveGraph and Teseo only support SI. In Section~\ref{sec-serializable-exp}, we use an additional experiment to compare the SI and Serializable versions of \sys{}.}
%Libin: the below discusses why Spruce was not compatible with other systems in our experiments.
%\textcolor{red}{%While Spruce~\cite{spruce-ref} also supports transactions, 
%Libin: I found their timestamp protocol is incorrect. I can add a sentence: "Its timestamp and snapshot protocol also do not guarantee transaction isolation.", "it cannot support transaction isolation." But should we also explain this? This can be fixed by all using the same global timestamp to fix this issue. But for the one adjacency list consistent version this is not an issue. Note its single operation transaction is still correct.
%We find 
Spruce~\cite{spruce-ref,spruce-code-ref} lacks enough transactional support for the experiments. 
Its transactions can only update a single directed edge, or read a single vertex's neighborhood. %The minimum requirement of our experiment is inserting an undirected edge atomically and 
To insert an undirected edge, Spruce creates 2 transactions (with two timestamps), breaking reciprocal consistency~\cite{chengmammoths,rec-consis-ref}. Also, it creates 1 transaction per adjacency list scan, thus, failing to obtain a transaction consistent view of the whole graph. Thus, we exclude this system from the experiments.
Table~\ref{tab:system-comparison} compares these systems. Note that Sortledton %supports serializable transactions but 
requires transactions to know their read and write sets beforehand~\cite{Sortledton}.
Experiments include a few instances of uniform graphs but the main focus is on power-law graphs as many real-world graphs exhibit such patterns~\cite{bank-power-law,powerlyra-ref,power-law-ref, taobench-ref}. Experiment results follow the color codes in Figure~\ref{fig:legends}.
%\Libin{todo: add a small figure about the metadata of each graph, I recall the reviewers mentioned this before}
%%\vspace{-6mm}

\begin{figure*}[htbp]
    \centering
    \includegraphics[width=1\textwidth]{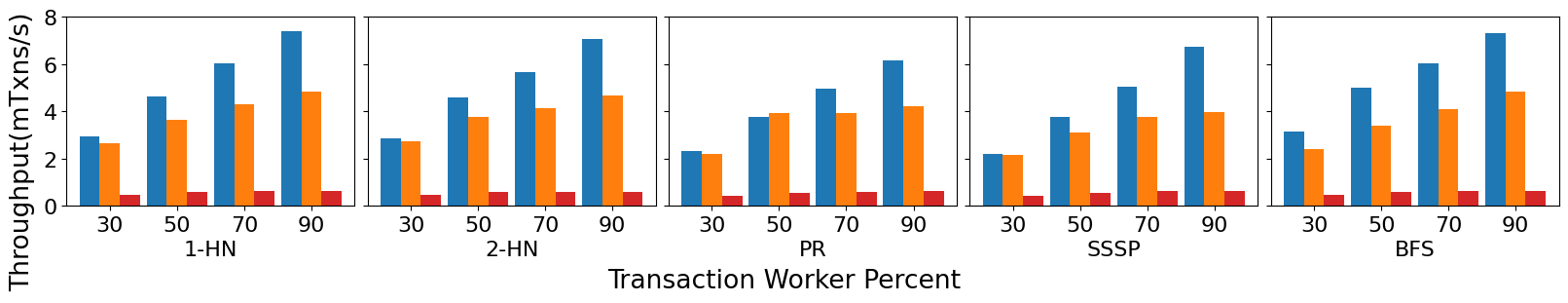}
    \caption{\textcolor{black}{Concurrent Updates and Analytics: Transaction Throughput}}
    %\vspace{-2mm}
    \label{fig:mixed-throughput}
\end{figure*}
\begin{figure*}[htbp]
    \centering
    \includegraphics[width=1\textwidth]{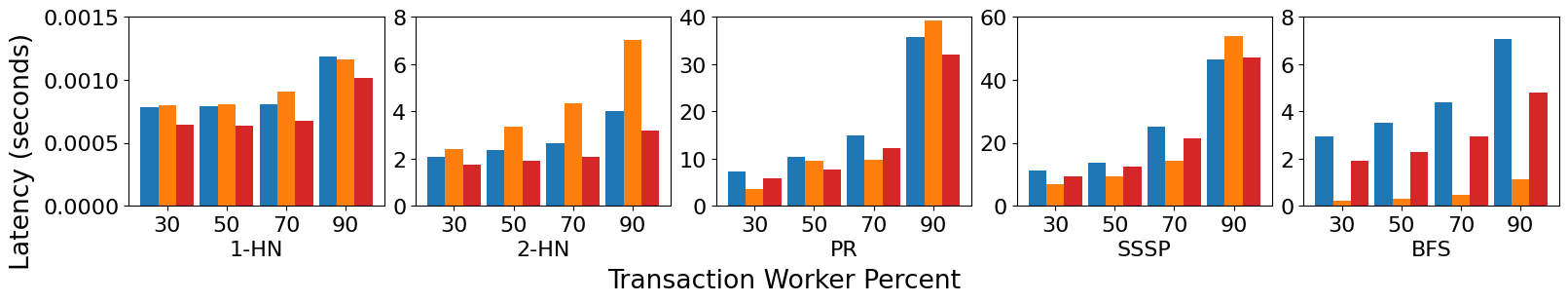}
    \caption{\textcolor{black}{Concurrent Updates and Analytics: Graph Analytics Latency}}
    %\vspace{-2mm}
    \label{fig:mixed-latency}
\end{figure*}
%\vspace{-2mm}
\subsection{Insert Performance}
%\Libin{March 28th: color in figures. Professor you told me to switch to filling instead of colors for figures, but I'm concerned that some figures have too many entries (so the bars are thin. It will be hard to add in fillings for them.)}
\label{sec-insertion-exp}
This experiment includes inserting uniform and power-law real-world and synthetic graphs with edges shuffled or according to timestamp order. %and inserting real-world power-law graphs according to edge timestamps. 
%We simulate undirected edge inserts by inserting 2 directed edges in a transaction.
Each system inserts an undirected edge by inserting 2 directed edges in a transaction 
and executes ``checked" operations so that every edge insert involves a read to check if the edge exists. Thus, each transaction internally executes 2 edge lookups and 2 edge writes atomically. Thus, having efficient edge reads can improve transaction throughput. %transactions' write performance. Figure~\ref{fig:legends} gives the color codes we use in displaying results. Refer to Figure~\ref{fig:insertion} for results.  %In these figures, the x-axis represents the number of worker threads and the y-axis represents throughput in the number of millions of transactions per second.\\

\noindent\textbf{Random-order Inserts}. %In this study, t
Evaluated systems insert edges from uniform and power-law real-world and synthetic graphs: {\em graph500-24, 26, uniform-24, dota-league}~\cite{gfe-dataset-ref,graphalytics-data-ref} and {\em twitter}~\cite{konect-ref}.  {\em graph500-26} and {\em twitter} have over 1 billion edges. Thus, we evaluate them  on the machine with larger DRAM up to 20 worker threads.
We have evaluated other %power-law and uniform 
graph datasets but their results are similar to the ones presented, and are omitted for brevity. The results are in Figure~\ref{fig:insertion}.
%Therefore we omit them here for space reason. \Libin{April 12th: not sure if we can still say we will include them in technical report} %but they can be found in our technical report.
%Bw-Graph is the second best overall system. 
%WGA: Why not report the experiments that Bw-Graphs perform better first, and then have these experiments?
%Libin: I will do that. Or maybe even exclude uniform graphs? Yes, at least in the first phase, because our paper, if not rejected, will have to go over another round of revision. We can put them back then.
%Let me reorder the paragraph first.
\sys{} 
%has the best scalability and outperforms all competitors for power-law graphs. It 
has over 10x better throughput than 
LiveGraph and 1.34x to 2x better throughput than the second-best Sortledton. 

\noindent{\bf Timestamp-ordered Inserts}. %In this experiment, 
We insert edges of real-world graphs {\em yahoo-songs} and {\em edit-wiki}~\cite{konect-ref} based on their creation timestamps. Real-life power-law graphs not only have hub vertices (dataset {\em edit-wiki} has vertices with millions of adjacent edges) but also hotspots.
%\wga{important. may move to intro sec.}
%\Libin{I thought we kind of already mentioned that in intro}
%: during a small duration many transactions need to concurrently update the same adjacency list, and those edge updates are considered to have strong temporal locality. 
%Previous work shows that all t
Transactional graph systems have significant performance degradation when inserting edges in timestamp order~\cite{Sortledton}. Thus, we compare performances %run the experiments 
%with the 4 systems 
in both random and timestamp order insertions. Refer to Figures~\ref{fig:insertion}'s {\em yahoo} and {\em edit-wiki}. As baseline, we evaluate inserting edges in random order. For {\em yahoo-songs}, \sys{} has the best scalability %and can achieve over 
with up to $70\%$ better throughput over Sortledton. %the second best Sortledton at higher parallelism degree. 
For {\em edit-wiki}, \sys{} ranks the second %overall ranks the second among the evaluated systems 
and performs the best with medium parallelism degree. %(1.2x better than Sortledton). 
\sys{} does not scale well here at high worker thread numbers because its adaptive delta chains adjustment heuristic does not catch up well with large neighborhoods. \sys{} could have used a more aggressive heuristic to allocate larger edge-deltas block and increase delta chain numbers during consolidation, so more concurrent writers would be allowed per edge-deltas block. However, this will increase the size of allocated blocks and memory used by indexes, and may incur wasted spaces for low-degree vertices. We plan to explore more sophisticated heuristics in future works. In contrast, when inserting in the timestamp order, \sys{} performs the best over all competitor systems. %\wga{by how much?}
%We have presented their results in log scale to accommodate low performing systems.
%Libin: Are these 2 sentences needed?
%Teseo experienced the worst throughput degradation which is much worse than what previous work reported~\cite{Sortledton}. We suspect the sparse array configuration may be different among the experiments.\\
The second-best Sortledton's throughput suffers up to $90\%$ degradation. %degrades by up to $90\%$. %to less than $11\%$ of its random order insertion throughput. 
In contrast, \sys's throughput only drops by up to $30\%$. In {\em yahoo-songs}, \sys{} has up to 11x better throughput than Sortledton and in {\em edit-wiki}, \sys{} has up to 9x better throughput. 

\noindent{\bf Performance Analysis}. Experiment results show that \sys{} has the highest throughput in power-law graph updates and only \sys{} can maintain high performance under edge update hotspot and temporal localities.  %\sys{} outperforms its competitors in most of the power-law graph experiments. 
We attribute \sys{}'s performance to its non-blocking latch-free graph store and low-overhead and adaptive delta-chain locking. \sys{'s} exclusive locks are embedded in the delta-chains index with minimal overhead. \sys{} adjusts the number of delta chains of each edge-deltas block as edge updates arrive, and enables concurrent writes to the same vertex. Thus \sys{} achieves the highest performance in timestamp-ordered workload. %due to its better concurrency provided by the adaptive delta chain locking. %While Teseo can reduce the conflict in updating the hub vertices by storing their neighborhoods into multiple segments, its sparse array segment-based storage has drawbacks. 
For uniform graphs, Sortledton outperforms \sys{} for low parallelism degrees, but \sys{}'s performance catches up as the number of concurrent worker threads increases. 
Since the graph is uniform, each vertex has about the same number of edges. Concurrent transactions are \textbf{not} likely updating the same adjacency lists. Thus, \sys{}'s %delta-chains index 
delta-chain locking
becomes under-utilized, and the drawbacks of vertex-centric locking are mitigated because transactions are less likely to conflict in this case. However, for power-law graphs,  Sortledton's threads block on concurrent updates on hub vertices and during a hotspot causing performance degradation. 
Ideally, Teseo should reduce the conflict in concurrently updating hub vertices by storing large neighborhoods in multiple segments. However, its sparse array segment rebalance exclusively locks multiple adjacent segments.
%but segments of the same neighborhood are always adjacent. 
Thus, concurrent rebalance threads may conflict when locking the same adjacent segment % in their rebalance window, or conflict 
or with other concurrent transactions. 
Teseo's segment locking may also incur false positives by locking vertices in the same segment. %among low-degree vertices by locking vertices in the same segment that the transaction does not intend to update. 
Thus, its throughput in timestamp-ordered experiment degrades by up to $99\%$.
%detailed in the context of the timestamp-ordered experiments.
%presented in Section{}. \wga{add sec. num. here.}
%may store multiple low degree vertices in the same segment and create false positive write-write conflicts. Moreover, its rebalance of segments, and split and merge of leaf nodes (sparse arrays) may exclusively lock neighborhoods of multiple vertices which degrades the update throughput. Such performance drawback is more significant in the next experiment. 
LiveGraph's performance suffers due to vertex-centric locking and the absence of edge indexes. For each edge insert, a LiveGraph transaction scans the edge block to check if the edge exists which is costly.
%Libin Oct 3rd: I remove the below 2 sentences
%LiveGraph, Sortledton, and Teseo suffer from temporal localities in edge update hotspot and power-law distributions. In contrast, \sys{} allows concurrent transactions to update the same vertex and adjusts the vertex's concurrency level adaptively. 

{\noindent \textbf{Lessons learned:}} Using latch-free techniques and low-overhead locking bits, having edge indexes, and breaking down vertex-centric locking to adapt workload enable higher transaction throughput.
\begin{figure*}[htbp]
    \centering
    \includegraphics[width=1\textwidth]{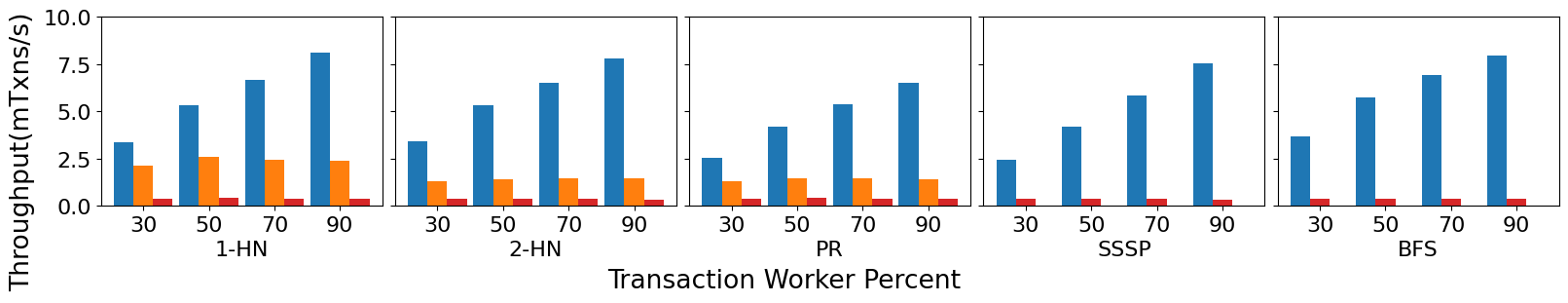}
    \caption{\textcolor{black}{Concurrent Update with Hotspot and Temporal Localities and Analytics: Transaction Throughput}}
    %\vspace{-2mm}
    \label{fig:mixed_hot_throughput}
\end{figure*}
\begin{figure*}[htbp]
    \centering
    \includegraphics[width=1\textwidth]{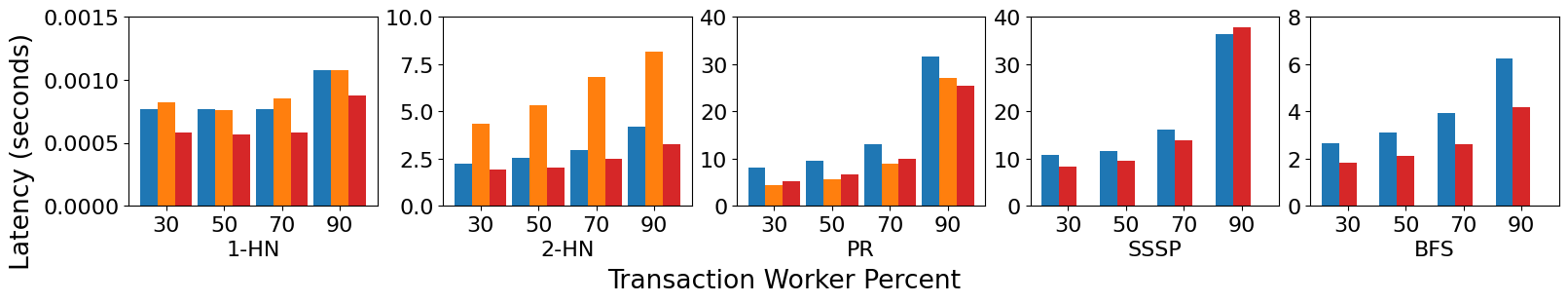}
    \caption{\textcolor{black}{Concurrent Update with Hotspot and Temporal Localities and Analytics: Graph Analytics Latency}}
    %\vspace{-2mm}
    \label{fig:mixed_hot_latency}
\end{figure*}
\subsection{\textcolor{black}{Update Performance}}
\textcolor{black}{We adopt the {\em graph500-24} update experiments from Teseo~\cite{Teseo} and Sortledton~\cite{Sortledton} to evaluate the performance of transactional graph systems in handling graph updates, but we add additional datasets.
We generate log files based on the edges of the power-law graph {\em graph500-24} using {\em graphlog}~\cite{graphlog-ref}. 
The source graph has around 260M  edges with 2.6B edge-update logs. The 1st 10\%  edge logs are edge inserts that build the original graph. The next 90\% are edge inserts and deletes. 
If a vertex has high degree in the  graph, it has proportionally more edge update logs. The experiment ensures that  graph size stays close to the initial size. All systems execute checked operations: Edge inserts (deletes) only happen after checking the edge existence. Each system uses one NUMA node. If memory is not enough, it uses a remote node.
LiveGraph's memory use is overly large, and could only process 20\% of the logs. 
Thus, its experiments are conducted using 20\% of the logs. We generate one set of edge update logs, and place the edge logs randomly in the edge log stream. %and the other having temporal localities and hotspots based on vertex IDs. 
Also, we  generate another set of edge logs with temporal localities and hotspots to mimic real-world scenarios. Transactions will more likely update the same vertex's adjacency list concurrently. 
We report the edge update transaction throughput in Figure~\ref{fig:update-throughput}.}

\textcolor{black}{\sys{} maintains its highest throughput and scalability in update transactions due to  reasons similar to  those discussed in the Insertion Experiments. Moreover, LiveGraph and Sortledton 
%similarly 
experience performance degradation in workloads with temporal localities and hotspots. One interesting difference is Teseo
%, who 
that 
maintains high performance in both workloads, and becomes the second-best performing system behind \sys{}. Notice that Teseo does not experience the same 
%catastrophic 
severe
performance degradation as 
that
in timestamp-ordered insertion experiment. We attribute this to the in-place updates in Teseo. As discussed earlier, the graph size stays close to its initial size during the update experiments, and Teseo performs in-place updates for deletes and subsequent inserts. Therefore, Teseo does not incur as many structure modifications as in the insert-only experiments. Service threads are invoked less frequently, and they are less likely to lock sparse array segments, reducing contentions among concurrent threads.}

{\noindent}\textcolor{black}{\textbf{Lessons Learned: } In-place updates reduce the frequency of structure modifications that have coarser lock-granularity, and may induce contention. Thus, limiting the amount of structure modifications in a graph system is beneficial for transaction throughput.}

\subsection{Concurrent Updates and Analytics}
\label{sec-mixed-workload}
%\Libin{March 10th: think about whether we call it update transaction or read-write transaction}
%\Libin{I feel like we have talked too much about the update only experiment, is it an issue? We omit its section in the paper}
%In this section, we evaluate
We study how systems support concurrent update transactions and graph analytics simultaneously. 

%We measure transaction throughput when running only updates. The results are similar to the transaction throughput results of this experiment and are omitted for brevity (One exception is Teseo's throughput increase due to its in-place updates and is the second best behind \sys). 
%\wga{to be considerate for the extra space in the paper, I am not using \\wga any more. Please see the embedded \%WGA99 that show in the tex document, but not in the pdf version.}
%WGA99: But does Teseo outperform GTX? You may need to say the throughput increase is by how much and that it is still lower than GTX.
%Libin: fixed
%\sys{} has up to 1.29x, 1.30x and 10.6x higher throughput than Sortledton, Teseo, and LiveGraph resp. Therefore we omit the details of the update throughput results here. %and include them in our technical report.
The mixed-workload experiment works as 
follows:
We allocate {\color{black}60} thread resources, and control the read/write ratio by assigning the threads either as write or OpenMP~\cite{openmp-ref} read threads, e.g., {\color{black} 30} write and {\color{black} 30} read threads form a 50\% write workload. Write threads continuously run update transactions \textcolor{black}{as in the Update Experiment using the same graph logs} while read threads run graph analytics. 
We 
%evaluate all 
%%previously mentioned 
%systems but 
could not report the results of Teseo as it runs into deadlocks (the same problem is reported in Sortledton's evaluation~\cite{Sortledton}). 
%Teseo can approximately achieve relatively good performance in BFS but its PageRank can be much worse than even Bw-Graph's PageRank. 
Sortledton has segmentation fault errors for some workloads, but we are still able to produce results. For analytics, we evaluate 1-hop (1-HN) and 2-hop neighbors (2-HN) that find 1-hop and 1-2 hop neighbors of a set of vertices, and breadth first search (BFS), PageRank (PR), and single source shortest paths (SSSP) from Graphalytics~\cite{graphalytics-paper,graphalytics-specification}. %\textcolor{red}{They cover the subgraph queries and global queries raised by Neo4j~\cite{aion-ref}.}
%Libin: Sept 29th: maybe omit the below sentence?
Similar experiments are in~\cite{Sortledton,spruce-ref} but they only run BFS and PR under certain thread configurations. %The results below  use the same color code in Figure~\ref{fig:legends}.
%are in Figures~\ref{fig:mixed-workload-normal-latency,fig:mixed-workload-normal-throughput,fig:mixed-workload-hotspot-latency,fig:mixed-workload-hotspot-throughput} 

%In this experiment, we evaluate concurrent graph topology scan (only scan the edges of the graph), graph property scan (scan the edges and their properties), BFS, and PageRank. All systems' BFS and PageRank are implemented according to~\cite{graphalytics-paper,graphalytics-specification}. For Bw-Graph, we enable prefetching of edge delta iterators if the workload is at least 30\% writes.\\
%\Libin{April 14th: Sortledton performs the best in BFS, much better than others with reasons given, is it legit to still discuss its performance but skip the result figures? Currently I don't feel like it is the right solution but to save space maybe? Or is there a way to gracefully acknowledge that they are better for BFS due to the reasons I explained but remove the figures? I feel we can save tons of space. But they fail for most of the BFS in hotspot workload though. Can we discuss how to present this? Can we say "they do extremly well in bfs because .... They maintain the same advantage in hotspot workload in bfs analytics but txn throughput degrades badly while the system crashes for most of the workloads?"}
%\textbf{Normal Update Distribution}. 
\noindent{\bf Normal Update Distribution}. \textcolor{black}{We use the same edge-update logs with random edge distribution as those in the Update Experiment.} %We generate edge update logs and place them randomly in the edge log stream. 
Write threads run transactions executing edge logs, and concurrent graph analytics start at 10\% of the update workload (after the source graph is loaded). The results are in Figures~\ref{fig:mixed-throughput}, \ref{fig:mixed-latency}. Sortledton crashes for several workloads. We redo these workloads 
%many times 
and record the results for the successful runs. %, but this was not always possible. Libin: Dec. 15th: under 60 threads, I sucessfully get Sortledton to run to all configurations for this experiment and have updated the figures.
%Note that some entries for Sortledton are missing because the system could not finish the experiment even once without crashing. 
%\Libin{I changed the below a bit because we are only reporting 30, 50, 70 and 90 percent now}
For 1-HN, PR, and SSSP, when the workload is read-heavy, e.g., 30\% write, \sys{}'s concurrent analytics can take at most $2\times$ longer time than the best performing {\color{black}competitor. %}Sortledton. 
%\sys{} and Sortledton also have similar transaction throughput under those workloads. 
%\metar{Meta.O2, R3.O2}{\color{red} 
\sys{}  has slightly higher transaction throughput than the second place Sortledton under these workloads.}
As the workload becomes  more write-heavy, \sys{}'s graph analytics performance catches up, and  has the highest transaction throughput. For write-heavy workloads, e.g., 90\% write, \sys{} has {\color{black}$1.46\times$ to $1.7\times$} %1.3x to 1.48x 
higher transaction throughput than the second best system Sortledton, and performs the best and second best in SSSP and PR, respectively. %Also, it has slightly 
{\color{black}\sys{} also maintains competitive performance in 1-HN against the other systems.} %worse performance in 1-HN than the other systems \metar{Meta.O2, R3.O2}{\color{red} depending on the workloads}. %However, \sys{} has better write throughput when executing BFS, but 
Sortledton wins in BFS with significantly lower latency under all workloads. 
\sys{} is {\color{black}$1.13\times$ to $1.75\times$} %1.12x to 1.52x 
faster than Sortledton in 2-HN under all workloads while achieving better transaction throughput {\color{black}($1.05\times$ to $1.51\times$).} %for most of the workloads (0.96x to \metar{Meta.O2, R3.O2}{\color{red} 1.7x}). %1.28x). 
%\textcolor{red}{\sys{} is  1.12x to 1.52x faster than Sortledton in 2-HN while achieving 1.01x to 1.28x higher transaction throughput under all workloads.}
LiveGraph exhibits good performance for graph analytics (around {\color{black}$0.99\times$ to $1.54\times$} %0.77x to 1.51x 
faster than \sys{}), but its transaction throughput is significantly lower than the other 2 systems (\sys{} has {\color{black}$5.2\times$ to $11.9\times$} %4x to 9x 
higher transaction throughput than LiveGraph). 
\begin{figure*}[htbp]
    \includegraphics[width=1\textwidth]{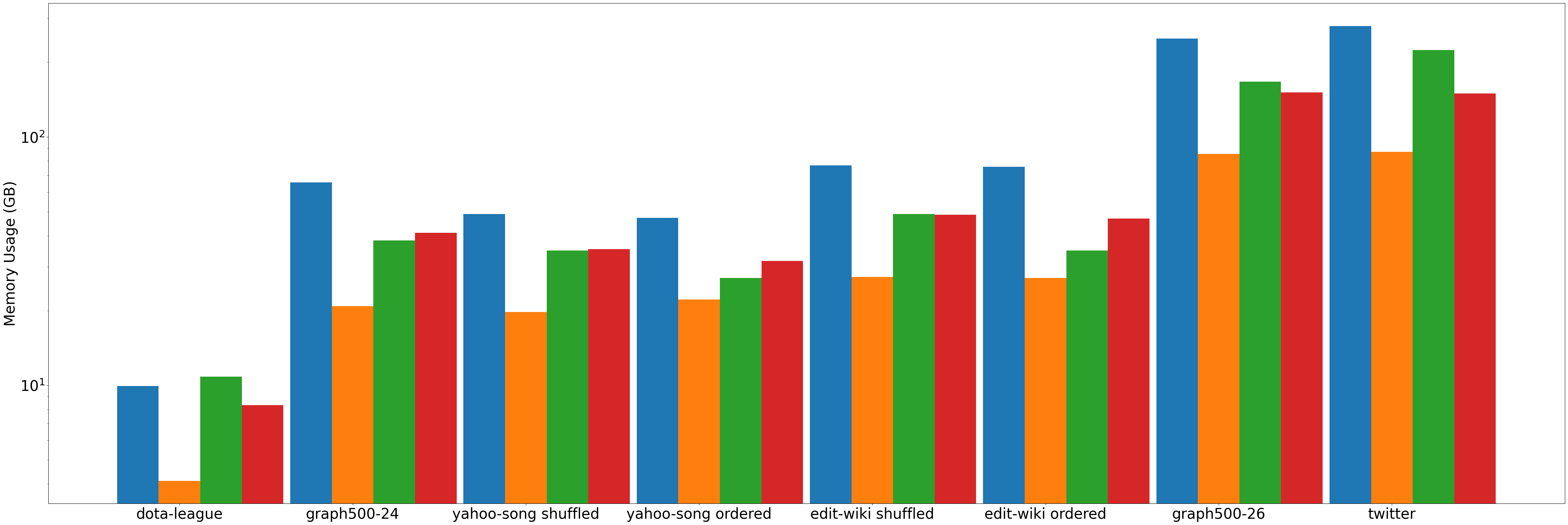}
    \caption{\textcolor{black}{Power-Law Graph Insertion Memory Usage}}
    \label{fig:memory-insert}
\end{figure*}
\begin{figure}[htbp]
     %\vspace{-3mm}
%\includegraphics[width=0.5\textwidth]{Figures/Matplotlib_Figures/mixed_workload_memory.png}
\includegraphics[width=0.5\textwidth]{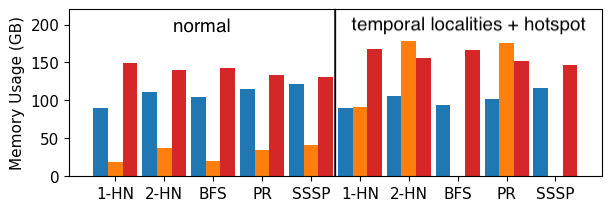}
     \caption{\textcolor{black}{Mixed-Workload Memory Consumption}}
     %\vspace{-1mm}
     \label{fig:memory}
\end{figure}
\noindent{\bf Temporal Locality and Hotspot Update Distribution}. 
%We generate a set of edge logs with temporal localities and hotspots to mimic real-world scenarios. Transactions will more likely update the same vertex's adjacency list concurrently. 
%Libin: sept 29th: I feel the next sentence can also be omitted. Its message was obvious.
This experiment has the same setting as the previous mixed-workload experiment except for having different edge logs. \textcolor{black}{We use the edge-update logs with temporal localities and hotspots from the Update Experiment.} We study performance for concurrent reads and writes with temporal localities. Sortledton's segmentation fault persists for this hotspots workload. 
%Sortledton's concurrent BFS and SSSP crashed even with multiple attempts. Therefore,
Thus, we only report available results in Figures~\ref{fig:mixed_hot_throughput}, \ref{fig:mixed_hot_latency}. 
%\Libin{April 12th: comment out the next paragraph}
%can also be found in our technical report. The results 
%are similar to the ones in insertion experiment with timestamps except that Teseo performs much better (the second best overall) than in the insertion experiment. \sys{} has up to 3.7x, 1.2x and 19.4x higher throughput than Sortledton, Teseo and LiveGraph respectively. The reasons behind Sortledton's and LiveGraph's performance degradation are similar to the ones in the timestamp-ordered insertion experiment. The in-place deletes in Teseo clear extra space for future insertions. Even with hotspots, Teseo invokes fewer segment rebalances and leaf nodes splits/merges. On the other hand, the timestamp-ordered inserts experiments involve only inserts. Segment rebalances and leaf node splits are unavoidable and performance drops significantly as a result. However, Teseo still cannot run in the following experiment. More details about the hotspot update experiments can be found in our technical report as well.
All systems' graph analytics performance are not affected much by update hotspots. 
%Each system limits the interference of transaction update patterns on concurrent graph analytics. 
%However, 
%transaction throughput is different from the results in the previous experiments. 
Sortledton and LiveGraph transactions 
degrade by up to {\color{black}68.5\% and 46\% }%64.3\% and 41.2\% 
in throughput, respectively,
for workloads with temporal localities and hotspots in update patterns.
%(up to 64.3\% and 41.2\% lower throughput for Sortledton and LiveGraph, respectively). 
%Meanwhile, 
Only \sys{} maintains high transaction throughput (up to {\color{black} $5.3\times$ and $22.3\times$ better than Sortledton and LiveGraph, respectively) } %$4.3\times$ better than Sortledton) 
and is competitive in graph analytics.
%WGA99: What do you mean by this last sentence? Does GTX also have lower throughput same as the other systems?
%Libin: fixed
%We conclude that their vertex-centric exclusive locking, and lack and adaptation to workload patterns and hotspots are the major causes of the performance degradation similar to their inferior throughput in the insertion experiments under temporal localities. \sys{}'s block and delta chains allocations are adaptive to the workload. Therefore its granularity of concurrency control (delta-chain level concurrency control) is adaptive to the hotspots and temporal localities. Therefore its performance is not negatively much by the update patterns. Instead, the hotspot and temporal locality update patterns are well-suited for \sys{}'s design and it even experiences improved throughput (up to 34\% higher throughput) and still maintains competitive analytics performance.
%Libin April 29th: end of commenting out
%Libin April 29th: adding new analysis subsection

\noindent{\bf Performance Analysis}. \sys{} has the best transaction throughput while maintaining competitive graph analytics performance across different workloads. We attribute this to the following reasons. 
\sys{} supports delta-based {\em MVCC} that prevents read transactions from conflicting with write transactions. Thus, the interference between concurrent updates and graph analytics is mitigated. 
%Thus, the side effects of concurrent updates are mitigated for graph analytics, and vice versa. 
%\sys{} outperforms vertex-centric locking and adapts to workload temporal localities and hotspots. 
\sys{'s} adaptive delta-chain locking adapts to workload temporal localities and hotspots to provide higher concurrency than vertex-centric locking systems.
Thus, \sys{} can maintain high transaction throughput across all experiments. 
It stores and manages edge delta-chains of each adjacency list sequentially in a memory block. This preserves cache locality, %eliminates pointer chasing, 
reduces random memory access and cache misses, and facilitates prefetch. Previous works~\cite{LiveGraph,Teseo,Sortledton} show that sequential adjacency list storage is beneficial to adjacency list scan: the backbone for graph analytics. 
\sys{} implements graph algorithms (e.g., BFS, PR, and SSSP) directly accessing edge-deltas blocks, and is compiled within the system. It saves the cost of using adjacency list iterators and associated functional calls. Our experiments
show that using an iterator is around $1.15\times$ slower in graph analytics. 
%Finally, \sys{} has several techniques in reducing the overhead of concurrent read-write transactions and graph analytics.
Finally, \sys{} distributes 
%Distributing 
the transaction table over worker threads to enable latch-free $O(1)$ access to transaction status for Lazy Updates, and the Block Access Table allows each reader and writer thread to register block-level access in its own cache-line-aligned entry without invalidating caches of other worker threads or modifying a shared lock.

Sortledton performs differently across experiments. It has the second best transaction throughput but it degrades for workloads with temporal localities and hotspots. 
For analytics, it performs the best in BFS. Sortledton uses exclusive lightweight latches for each adjacency list. BFS touches a smaller portion of the graph. Thus, update and read transactions are less likely to conflict, bypassing the negative locking effects. BFS requires each vertex's degree to initialize the algorithm. \sys{} and LiveGraph scan the vertex adjacency list to calculate this value. This dominates the BFS latency. Sortledton supports vertex degree versioning that allows transactions to get vertex degree efficiently, thus supporting efficient BFS. PR and SSSP require finding each vertex's degree, but they also need to scan vertices' adjacency lists multiple times. The algorithm's execution phase dominates the total latency, and thus Sortledton's advantages drop. As read threads scan the graph heavily, they become more likely to conflict with write transactions. %, and hinder the performance. 
1-HN and 2-HN only requires adjacency list scans. In 2-HN, each vertex's 1-hop neighbors are not known a priori, causing Sortledton to have large locking overhead, and hence has {\color{black}worse performance than \sys{}.} %similar or worse performance. 
LiveGraph shows strong disparity between its transaction throughput and graph analytics. Its transaction throughput is several times worse than both \sys{} and Sortledton but its graph analytics perform better. %better than other systems. 
LiveGraph uses a similar-styled $MVCC$ as \sys{}'s but only optimizes for sequential adjacency list scans~\cite{LiveGraph}. %It provides the same level of guarantees 
It guarantees that each vertex's adjacency list is stored sequentially in memory and transaction reads and writes never conflict. Moreover, LiveGraph has smaller edge log entries compared to \sys{'s} edge deltas. The compact sequential adjacency list storage enables good read performance. But LiveGraph suffers from its simple vertex-centric locking and lack of edge lookup indexes as discussed in Section~\ref{sec-insertion-exp} and has the worst transaction throughput. 

{\noindent \textbf{Lessons learned:}} Limiting writes to shared memory variables, isolating updates in versions, and being cache-friendly reduce transaction overheads and interference. Having a sequential adjacency list and limiting function calls enable performant graph analytics. 
\subsection{Memory Consumption}
\label{sec-memory}

%\Libin{March 24th: professor, please let me know if you think using a table will be better.}

%\iffalse
%\begin{figure}
%  \includegraphics[width=\linewidth]{Figures/GTX Updated Experiment/memory_usage.png}
%  \caption{Memory Usage in GB's}
%  \label{fig:memory}
%\end{figure}
%\fi
%\Libin{April 13th: remove the figure of insertion only. We focus only on the mixed workload because we target them and want to show our robust memory usage}
%We evaluate the total memory consumption of the various systems for different workloads. We use the memory usage per process stored by the OS to evaluate the memory consumed by each system for each experiment. 
{\noindent}\sys{} uses 64-byte edge deltas, so 
it uses more memory to load and store a graph. 
For %most graph inserts, 
power-law graph inserts, 
\sys{} uses around $2\times$ to $3\times$, $0.9\times$ to $2.1\times$, and $1.2\times$ to $1.9\times$ more memory than Sortledton, Teseo and LiveGraph respectively (\textcolor{black}{Figure~\ref{fig:memory-insert}}). %However, for workload with updates, concurrent analytics, and hotspots, \sys{} does not incur much more memory overhead than storing a static graph (pure inserts), and  memory consumption stays stable across workloads (Figure~\ref{fig:memory}). 
However, \sys{'s} memory usage stays stable and consistent in more complex workloads involving updates (invalidated versions), concurrent analytics, and hotspots. Figure~\ref{fig:memory} shows memory usage comparison of different systems under \textcolor{black}{balanced} mixed-workload experiments %\textcolor{red}{(under balanced read-write distribution with 30 read threads and 30 write threads)}
. %Libin: Dec.14th: this new sentence was not asked by the reviewers but I feel it can help clarify the data. Or we can omit it and leave it in the experiment.
It shows that \sys{} is efficient in memory allocation and garbage collection. LiveGraph uses more memory if the workload involves updates and analytics, while Sortledton's memory usage increases at most {\color{black}$5.17\times$} %6x 
in workloads with temporal localities and hotspots. \sys{} has the lowest memory consumption for these experiments. 
\begin{figure}[htbp]
    \centering
    \includegraphics[width=0.5\textwidth]{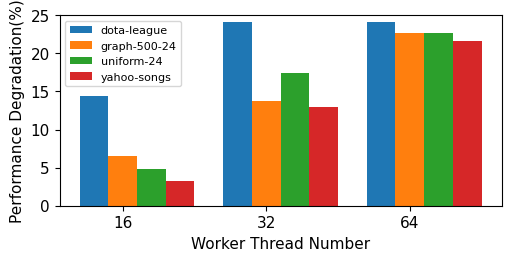}
    %\vspace{-6mm}
    \caption{\textcolor{black}{Performance Degradation of Serializable \sys{}}}
    \label{fig:si-serializable}
    %\vspace{-2mm}
\end{figure}
\subsection{\textcolor{black}{Serializable vs. Snapshot Isolation \sys{}}}
\label{sec-serializable-exp}
{\color{black}We modify the insertion experiment (Section~\ref{sec-insertion-exp}) to evaluate Serializable \sys{'s} overhead over its Snapshot Isolation (SI) version using the same datasets and number of threads. Besides the checked edge insertion, each worker thread records the edge inserted in its last transaction. For each subsequent edge insert, the worker thread creates a new read-write transaction that first reads the previous edge inserted, and then inserts the checked edge. %This experiment ensures each transaction contains a separate read set to validate. %A new experiment is needed because in our original experiments a read-write transaction does not contain a separate read set to validate. 
%We measure \sys{'s} performance using the same datasets and numbers of threads as in Section~\ref{sec-insertion-exp} and report the average degradation with 4 representative datasets in Figure~\ref{fig:si-serializable}. %We show the results of 4 representative datasets in Figure~\ref{fig:si-serializable}.
We report the average degradation using 4 of the datasets as in Figure~\ref{fig:si-serializable}. %Libin Jan. 8th: do you think the 4 datasets will be an issue here? Like we did not evaluate ALL datasets but only 4 of them. So that's why I said representative dataset.
Serializable \sys{} has 3-25$\%$ throughput degradation vs. the SI version due to the following factors. %The performance degradation is caused by the following factors. 
%The read set validation phase requires committing transactions to synchronize with the commit manager when the whole commit group validates their read sets. 
Committing transactions and the commit manager need to synchronize in the quiescent state to enforce serialization between commit groups.
Moreover, transactions need to recheck objects they read and will abort if they break serializability.}
%Libin Dec 17th: 2 concerns. 1 is that do you want me to evaluate even more datasets? Or 4 is enough? Another is that I did not talk about read set valdiation requires more synchronization with the commit manager (I feel this is implementation detail of the validation phase) when I discuss serializable support in section 6. Do you think here we talk about the need of synchronization in read set validation is weird and unnatural? Thanks

{\noindent\textbf{Lessons Learned:}} Serializable transaction support comes at a price of up to 25\% performance degradation. %It justifies that 
%State-of-the-art graph systems 
LiveGraph~\cite{LiveGraph} and Teseo~\cite{Teseo} only support SI. Sortledton~\cite{Sortledton} supports serializability but requires known read and write sets and finishing all reads before writes. Because serializability support is costly and SI is sufficient for graph workloads (discussed in Section~\ref{sec-txn-cc}), serializable transactions should be used only when necessary.
\section{Conclusions}
\label{bwgraph-conclusion}
\sys{} is a latch-free write-optimized transactional graph system that supports concurrent read-write transactions and graph analytics. It adopts an adjacency list format %similar to that of LiveGraph~\cite{LiveGraph} 
and a delta chain-based storage with delta-chains index. Its adjacency list design combines both the %traditional 
linked list-based delta store~\cite{BwTree} and append-only delta updates~\cite{openBwTree, LiveGraph} to enable high update throughput, low-latency single edge lookup, and sequential adjacency list scan. \sys{} has high read and write concurrency using transactions, multi-versioning, and Lazy Update hybrid commit under snapshot isolation. Evaluation against state-of-the-art transactional graph systems shows that \sys{} has better performance for read-write transactions, and  is several times better in handling real-world write workload with temporal localities. 
%Bw-Graph also handles the transactional read-write mixed workload well as our delta-chains index significantly speed up checked-writes and single edge reads. 
%We show while our graph analytics performance is not ideal in static environment, 
\sys{} performs well under mixed workloads when the workload is at least 50\% writes, in which case, \sys{} has better read-write transaction throughput while its graph analytics latency remains competitive. 
For most of the write-heavy workloads, 
\sys{} significantly outperforms competitors in write throughput while maintaining close or even better graph analytics latency. 
%\sys{} ties or outperforms competitors in graph analytics while significantly outperforming them in write throughput. 
%\sys{} is also one of the only 2 transactional graph systems that can complete all experiments without bugs and anomalies. 
Overall, \sys{} demonstrates that using latch-free techniques, avoiding vertex-centric locking, being cache-friendly, maintaining adjacency list-level edge indexes and sequential adjacency list storage, and being adaptive to the workload can enable high transaction throughput and better read-write concurrency in transactional graph systems.
\bibliographystyle{ACM-Reference-Format}
\bibliography{sample-base}

\appendix

\end{document}